\documentclass[11pt,showpacs,superscriptaddress]{revtex4}
\usepackage{bm}
\usepackage{dcolumn}
\usepackage{indentfirst}
\usepackage{longtable}
\usepackage{graphicx,epsfig,latexsym,amssymb}
\usepackage{multirow,amsmath,array,booktabs,color}
\usepackage[section]{placeins}

\newcommand{\beq}{\begin{equation}}
\newcommand{\eeq}{\end{equation}}
\newcommand{\beqn}{\begin{eqnarray}}
\newcommand{\eeqn}{\end{eqnarray}}
\newcommand{\bea}{\begin{array}}
\newcommand{\eea}{\end{array}}
\newcommand{\bsub}{\begin{subequations}}
\newcommand{\esub}{\end{subequations}}
\newcommand{\ff}[1]{\frac{1}{#1}}

\newcommand{\rc}{\right>}
\newcommand{\lr}{\left|}

\newcommand{\lrb}[1]{\left(#1\right)}
\newcommand{\lrs}[1]{\left[#1\right]}

\newcommand{\svec}[1]{\boldsymbol{#1}}
\newcommand{\ivec}{\vec}

\begin{document}

\title{Neutron star properties in density dependent relativistic Hartree-Fock theory}

\author{Bao Yuan Sun}
 \affiliation{School of Phyics£¬and State Key Laboratory of Nuclear Physics and Technology, Peking
University, 100871 Beijing, China}
\author{Wen Hui Long}
 \affiliation{School of Phyics£¬and State Key Laboratory of Nuclear Physics and Technology, Peking
University, 100871 Beijing, China}
 \affiliation{Physik-Department der Technischen Universit\"at M\"unchen, D-85748 Garching, Germany}
\author{Jie Meng}
 \affiliation{School of Phyics£¬and State Key Laboratory of Nuclear Physics and Technology, Peking
University, 100871 Beijing, China}
 \affiliation{Department of Physics, University of Stellenbosch, Stellenbosch, South Africa}
 \affiliation{Institute of Theoretical Physics, Chinese Academy of Sciences, 100080 Beijing, China}
 \affiliation{Center of Theoretical Nuclear Physics, National Laboratory of Heavy Ion Accelerator,
730000 Lanzhou, China}
\author{U. Lombardo}
 \affiliation{Laboratori Nazionali del Sud, Istituto Nazionale di Fisica Nucleare, Via S. Sofia 62,
 I-95123 Catania, Italy}
 \affiliation{Dipartimento di Fisica e Astronomia, Universit\'{a} di Catania, viale Andrea Doria 6,
 I-95125 Catania, Italy}

\begin{abstract}
With the equations of state provided by the newly developed density
dependent relativistic Hartree-Fock (DDRHF) theory for hadronic
matter, the properties of the static and $\beta$-equilibrium neutron
stars without hyperons are studied for the first time, and compared
to the predictions of the relativistic mean field (RMF) models and
recent observational data. The influences of Fock terms on
properties of asymmetric nuclear matter at high densities are
discussed in details. Because of the significant contributions from
the $\sigma$- and $\omega$-exchange terms to the symmetry energy,
large proton fractions in neutron stars are predicted by the DDRHF
calculations, which strongly affect the cooling process of the star.
The critical mass about 1.45 $M_\odot$, close to the limit 1.5
$M_\odot$ determined by the modern soft X-ray data analysis, is
obtained by DDRHF with the effective interactions PKO2 and PKO3 for
the occurrence of direct Urca process in neutron stars. The maximum
masses of neutron stars given by the DDRHF calculations lie between
2.45 M$_\odot$ and 2.49 M$_\odot$, which are in reasonable agreement
with high pulsar mass $2.08 \pm 0.19~M_\odot$ from PSR B1516+02B. It
is also found that the mass-radius relations of neutron stars
determined by DDRHF are consistent with the observational data from
thermal radiation measurement in the isolated neutron star RX J1856,
QPOs frequency limits in LMXBs 4U 0614+09 and 4U 1636-536, and
redshift determined in LMXBs EXO 0748-676.
\end{abstract}

%
%
%
%

\maketitle

\section{Introduction}
The investigative territory of nuclear physics has been enormously
expanded with the construction of the new accelerator facilities as
well as the development of the land- and space-based observatories.
The exploration over the phase diagram of matter has been extended
to the extreme conditions of density, pressure and temperature
during the last several decades, which is now one of the hottest
topics in both theoretical and experimental nuclear physics. While
around the saturation density, the nuclear matter properties can be
well calibrated by terrestrial experiments with the atomic nuclei,
the neutron stars are the natural laboratories in the unverse for
exploring the equation of state (EoS) of baryonic matter at low
temperature and higher baryonic densities \cite{BookWeber99,
Shapiro83, Weber:2007}. In addition, the probe to the elliptical
flow and kaon production in heavy-ion collisions provides extra
information for the nuclear EoS at high temperature and about
$2\sim4.5$ times nuclear saturation density~\cite{Danielewicz:2002,
Fuchs:2006Reports, BALi:2008Reports}.

As one of the most exotic objects in the universe, neutron star
plays the role of a bridge between nuclear physics and astrophysics.
With the discovery of neutron in 1932, the concept of the neutron
star were firstly proposed by Landau \cite{Landau:1932}. Two years
later, the neutron star was deemed to be formed in \emph{supernovae}
\cite{Baade:1934}. In 1960s, the observed radio pulsars
\cite{Hewish:1968} were identified as the rotating neutron stars
\cite{Pacini:1968}. Currently the neutron star is generally
considered to be of the crust structure \cite{Lattimer:2004}. Below
the atmosphere and envelope surface with a negligible amount of
mass, the crust extends about 1 to 2 km into star, which mainly
consists of nuclei and free electrons. With the density increasing,
the dominant nuclei in the crust vary from $^{56}$Fe to extremely
neutron-rich nuclei and neutrons may gradually leak out of nuclei to
form the neutron fluid. The outer core ($\rho\gtrsim\rho_0/3$) of
neutron stars is composed of a soup of nucleons, electrons, and
muons. In the inner core, exotic particles may become abundant, such
as the strangeness-bearing hyperons and/or Bose condensates (pions
or kaons), and a transition to a mixed phase of hadronic and
deconfined quark matter becomes possible. Although similar EoS at
saturation and subsaturation densities is obtained by various
nuclear matter models, their deviations are very remarkable in the
high density region, which is very essential in describing and
predicting the properties of neutron stars. Further investigations
are therefore necessitated for the detailed structure over the
density range of neutron stars.

The recent observations of neutron stars have been reviewed, such as
in Ref. \cite{Lattimer:2007}. The existence of massive compact stars
of 2 $M_\odot$ or above is now unveiled by some evidence. Careful
analysis of the \emph{Rossi X-ray Timing Explorer} (\emph{RXTE})
data for the quasi-periodic brightness oscillations (QPOs)
discovered from low-mass X-ray binaries (LMXBs) 4U 1636-536 shows
that several neutron stars in LMXBs have the gravitational masses
between 1.9 $M_\odot$ and possibly 2.1 $M_\odot$
\cite{Barret:2005wd}. Measurements on millisecond pulsars in
globular cluster NGC 5904 (M5) during 19 years of Arecibo timing
yield $M=2.08 \pm 0.19~M_\odot$ for PSR B1516+02B
\cite{Freire:2007}. Whereas a much larger pulsars mass $2.74 \pm
0.21~M_\odot$ is presented very recently for PSR J1748-2021B in NGC
6440 \cite{Freire:2008}. Besides the maximum mass limits, the
mass-radius relation is also constrained by the recent observations.
The thermal radiation spectra in X-rays and in optical-UV from the
isolated neutron star RX J1856.5-3754 (shorthand: RX J1856)
determine the large radiation radius $R_\infty$ as 16.8 km
\cite{Trumper:2003we}. The model fitting to the high-quality X-ray
spectrum of the quiescent LMXB X7 in the globular cluster 47 Tuc
prefers a rather large radius of $14.5^{+1.8}_{-1.6}$ km for a 1.4
$M_\odot$ compact star \cite{Heinke:2006}. In another LMXB, EXO
0748-676, a pair of resonance scattering lines consistent with Fe
XXV and XXVI, gives the redshift $z$ about $0.345$, which constrains
the mass $M\ge 2.10\pm 0.28~M_\odot$ and the radius $R \ge 13.8 \pm
1.8$ km for the same object \cite{Cottam:2002, Ozel:2006km}. In
addition, the highest QPOs frequency 1330 Hz ever observed in 4U
0614+09 implies the mass $M\lesssim1.8~M_\odot$ and the radius
$R\lesssim 15$ km in this object \cite{Miller04}. Furthermore,
modern observational soft X-ray data of cooling neutron stars
associated with popular synthesis models' analysis reveal that an
acceptable EoS shall not allow the direct Urca process
\cite{Lattimer:1991} to occur in neutron stars with masses below 1.5
$M_\odot$ \cite{Blaschke:2004, Popov:2006, Klahn:2006ir}. All of
these indicate the strict constraints on the EoS of strongly
interacting matter at high densities.

For the description of nuclear matter and finite nuclei, the
relativistic many-body theory has achieved great success during the
past years. One of the most successful representatives is the
relativistic Hartree approach with the no-sea approximation, namely
the relativistic mean field (RMF) theory \cite{Miller:1972,
Walecka:1974, Serot:1986}. With a limited number of free parameters
including the meson masses and meson-nucleon coupling constants, the
appropriate quantitative descriptions are obtained by RMF for both
stable nuclei and exotic ones with large neutron excess
\cite{Reinhard:1989, Ring:1996, Serot:1997, Bender:2003, meng98npa,
meng96prl, meng98prl, meng98plb, Meng:2006, Lalazissis:1997,
Typel:1999, Long04, Niksic:2002, Lalazissis:2005}.

After the first theoretical calculations of the neutron star
properties \cite{Oppenheimer:1939, Tolman:1939}, a plenty of
theoretical prediction was made by both non-relativistic and
relativistic approaches in the literature. At the early development
of RMF, it was applied to evaluate the total mass and radius of
neutron stars \cite{gle}. In the further development, the nuclear
medium effects were taken into account by introducing the explicit
or implicit density dependence into the meson-nucleon couplings,
i.e., the density dependent meson-nucleon couplings
\cite{Brockmann:1992, Lenske:1995, Fuchs:1995} and the non-linear
self-couplings of the meson fields \cite{Boguta:1977, Sugahara:1994,
Long04}, respectively. In Refs. \cite{Boguta:1977, Serot:1979,
Sumiyoshi:1995}, the effects of the non-linear self-coupling of
$\sigma$, $\omega$ and $\rho$ mesons were studied in describing the
nuclear matter and neutron stars. On the other side, the influence
on mean field potentials, saturation properties of nuclear matter,
the EoS, the maximum mass and radius of neutron stars was
systematically investigated with explicit density dependence in the
meson-nucleon couplings \cite{Hofmann:2001, Ban:2004}. In addition,
the consequences on compact star properties were studied with the
inclusion of the degree of freedom of hyperons
\cite{Glendenning:1982, gle85, Knorren:1995, Schaffner:1996}. With
more accurate experimental data of the neutron radius of $^{208}$Pb,
the correlation between the neutron skin thickness in finite nuclei
and the symmetry energy of nuclear matter was discussed
\cite{Horowitz:2001, Horowitz:2001PRC, Brown:2007}. Further
investigation in Ref. \cite{Horowitz:2002} proposed the relation
between the neutron skin thickness of a heavy nucleus and neutron
star radius that the larger neutron skin thickness prefers the
larger neutron star radius, which implies the constraints on the EoS
and on the cooling mechanism of neutron stars. Besides the RMF
approach, the Dirac-Brueckner-Hartree-Fock (DBHF) and
Brueckner-Hartree-Fock (BHF) with three-body force approaches were
also applied to study the neutron star properties with the realistic
nucleon-nucleon interactions \cite{Huber:1994, Engvik:1994,
Krastev:2006, BHF1, BHF2}.

In the RMF approach, however, the Fock terms are neglected by the
reason of simplicity. From the recent development on the
relativistic Hartree-Fock theory, i.e., the density dependent
relativistic Hartree-Fock (DDRHF) theory \cite{Long:2006}, it is
found that the Fock terms are of special importance in determining
the nuclear structure properties. Within DDRHF, the quantitatively
comparable precision with RMF are obtained for the structure
properties of nuclear matter and finite nuclei \cite{Long:2006,
Long:2007}. Particularly, the new constituents introduced with the
Fock terms, i.e., the $\rho$-tensor correlations and pion exchange
potential, have brought significant improvement on the descriptions
of the nuclear shell structures \cite{Long:2007} and the evolutions
\cite{Long:2008}. Furthermore, the excitation properties and the
non-energy weighted sum rules of the Gamow-Teller resonance and the
spin-dipole resonance in the doubly magic nuclei have been well
reproduced by RPA based on the DDRHF approach fully
self-consistently \cite{Liang:2008}. Since the nuclear structure
properties around the saturation density are evidently affected by
the Fock terms, one might expect remarkable effects from the Fock
terms on the nuclear matter properties in the high density region.
Especially with the inclusion of the new ingredients in DDRHF,
remarkable adjustment occurs on the coupling strength of the
dominant mean fields ($g_\sigma$ and $g_\omega$), which may bring
significant effects when exploring to the high density region.

In this paper,  the properties of the static and $\beta$-equilibrium
neutron stars without hyperons are studied within the DDRHF theory.
As compared to the calculations of the RMF theory, the applicable
ranges of density and isospin asymmetry are tested for DDRHF as well
as the consistence with recent observational constraints of compact
stars. In Section \ref{sec:GFNM} is briefly introduced the formulism
of DDRHF for nuclear matter and neutron star. In Section
\ref{sec:RD}, the calculated results and discussions are given,
including the properties of symmetric and asymmetric nuclear matter
in comparison with RMF in Section \ref{sec:RDNM}, in which the
effects of Fock terms are studied in details, and the investigations
of neutron stars in comparison with recent observational data in
Section \ref{sec:RDNS}. Finally a summary is given.

\section{General Formulism of DDRHF in Nuclear Matter}\label{sec:GFNM}
The relativistic Hartree-Fock (RHF) theory with density dependent
meson-nucleon couplings, i.e., the DDRHF theory, was firstly
introduced in Ref. \cite{Long:2006}, and the applications and
corresponding effective interactions can be found in Refs.
\cite{Long:2006, Long:2006PS, Long:2007, Long:2008}. In the
following we just briefly recall the general formulism of DDRHF in
nuclear matter and the application in neutron stars. More details of
the RHF theory are referred to Refs. \cite{Bouyssy:1987, Long:2006,
Long:2007}.

As the theoretical starting point, the Lagrangian density of DDRHF is constructed on the one-boson
exchange diagram of the NN interaction, which contains the degrees of freedom associated with the
nucleon ($\psi$), two isoscalar mesons ($\sigma$ and $\omega$), two isovector mesons ($\pi$ and
$\rho$), and the photon ($A$). Following the standard procedure in Ref. \cite{Bouyssy:1987}, one
can derive the Hamiltonian in nucleon space as,
 \begin{equation}\label{Hamiltonian in nucleon space}
H=\int d^3 x\ \bar\psi\lrb{-i\svec\gamma\cdot\svec\nabla + M}\psi +\frac{1}{2} \int d^3x d^4y
\sum_{\phi} \bar\psi(x) \bar\psi(y) \Gamma_\phi(1,2)D_\phi(1,2) \psi(y)\psi(x),
 \end{equation}
where $\phi=\sigma, \omega, \rho, \pi$ and $A$, and $D_\phi$ denotes
the propagators of mesons and photon. The interacting vertex
$\Gamma_\phi$ in the Hamiltonian (\ref{Hamiltonian in nucleon
space}) reads as,
 \begin{subequations}\label{interaction matrix}
\begin{eqnarray}
    \Gamma_\sigma(1,2) &\equiv& -g_\sigma(1)g_\sigma(2),\\
    \Gamma_\omega(1,2) &\equiv& +g_\omega(1)\gamma_\mu(1)g_\omega(2)\gamma^\mu(2),\\
    \Gamma_\rho(1,2) &\equiv& +g_\rho(1)\gamma_\mu(1)\vec\tau(1)\cdot g_\rho(2)\gamma^\mu(2)\vec\tau(2),\\
    \Gamma_\pi(1,2) &\equiv& -\ff{m_\pi^2}\lrs{f_\pi\ivec\tau\gamma_5\svec\gamma\cdot\svec\nabla}_1
        \cdot\lrs{f_\pi\ivec\tau\gamma_5\svec\gamma\cdot\svec\nabla}_2,\\
    \Gamma_A(1,2) &\equiv& +\frac{e^2}{4}[\gamma_\mu(1-\tau_3)]_1[\gamma^\mu(1-\tau_3)]_2.
\end{eqnarray}\end{subequations}
In current work, the $\rho$-tensor correlations are not enclosed. In the above expressions and
following context, the isovectors are denoted by arrows and the space vectors are in bold type.

In general, the time component of the four-momentum carried by
mesons are neglected on the level of the mean field approximation.
This neglect has no consequence on the direct (Hartree) terms, while
for the exchange (Fock) terms it amounts to neglect the retardation
effects. The meson propagators are therefore of the Yukawa form,
e.g., in the momentum representation,
 \begin{equation}\label{Yukawa propagator}
D_\phi(1,2) = \frac{1}{m_\phi^2 + \svec q^2},
 \end{equation}
where the exchanging momentum $\svec q = \svec p_2 - \svec p_1$, and $\phi = \sigma, \omega, \rho$
and $\pi$.

For the description of nuclear matter, the coulomb field thus could be neglected, and the momentum
representation is generally adopted in the Hamiltonian. Due to time-reversal symmetry and
rotational invariance, the self-energy $\Sigma$ can be expressed as
\begin{equation}\label{Total Self Energy}
\Sigma(p)=\Sigma_S(p)+\gamma_0\Sigma_0(p)+\boldsymbol{\gamma}\cdot\hat{\boldsymbol{p}}\Sigma_V(p),
\end{equation}
where $\hat{\svec p}$ is the unit vector along $\svec p$, and the scalar component $\Sigma_S$, time
component $\Sigma_0$ and space component $\Sigma_V$ of the vector potential are functions of the
four-momentum $p=(E(p), \svec p)$ of nucleon. With the general form of the self-energy, the Dirac
equation in nuclear matter can be written as,
 \begin{equation}\label{Dirac equation in momentum representation}
\lrb{\svec\gamma\cdot\svec p^* + M^*} u(p,s,\tau)=\gamma_0E^*u(p,s,\tau),
 \end{equation}
with the starred quantities,
 \bsub\label{Starred}\begin{eqnarray}
\boldsymbol{p}^* &=& \boldsymbol{p}+\hat{\boldsymbol{p}}\Sigma_V(p),\\
M^* &=& M+\Sigma_S(p),\\
E^* &=& E(p)-\Sigma_0(p),
 \end{eqnarray}\esub
which obey the relativistic mass-energy relation ${E^*}^2=\svec p^{*2}+M^{*2}$. With this
relationship, one can introduce the hatted quantities as
 \begin{align}\label{hatted quantities}
    \hat P\equiv&\frac{\boldsymbol{p}^*}{E^*},& \hat M\equiv&\frac{M^*}{E^*}.
 \end{align}
With the momentum representation, the Dirac equation (\ref{Dirac equation in momentum
representation}) can be formally solved and the Dirac spinors with positive energy read as
 \begin{equation}\label{positive energy solutions}
    u(p,s,\tau)=\left[\frac{E^*+M^*}{2E^*}\right]^{1/2}
        \left(\begin{array}{c}
            1 \\ \displaystyle\frac{\boldsymbol{\sigma}\cdot\boldsymbol{p}^*}{E^*+M^*}
        \end{array}\right)
        \chi_s \chi_\tau,
 \end{equation}
where $\chi_s$ and $\chi_\tau$ respectively denote the spin and isospin wave functions. The
solution (\ref{positive energy solutions}) is normalized as
\begin{equation}
    u^\dag(p,s,\tau)u(p,s,\tau)=1.
\end{equation}

The stationary solutions of the Dirac equation (\ref{Dirac equation
in momentum representation}) consist of  the positive and negative
energy ones, and one can expand the nucleon field operator $\psi$ in
terms of Dirac spinors. Within the mean field approximation, the
contributions from the negative energy states are neglected, i.e.,
the no-sea approximation. The nucleon field operate $\psi$ is
therefore expanded on the positive energy set as,
 \bsub\begin{align}
    \psi     (x)=&\sum_{p,s,\tau}u(p,s,\tau)e^{-ipx}c_{p,s,\tau}\ ,\\
    \psi^\dag(x)=&\sum_{p,s,\tau}u^\dagger(p,s,\tau)e^{ipx}c^\dagger_{p,s,\tau}\ .
\end{align}\esub
where $c_{p,s,\tau}$ and $c^\dagger_{p,s,\tau}$ are the annihilation and creation operators. With
the no-sea approximation, the trial Hartree-Fock ground state can be constructed as,
\begin{equation}\label{trial ground state in momentum representation}
    \left| \Phi_0 \right>=\prod_{p,s,\tau}c^\dag_{p,s,\tau}\left| 0\right>,
\end{equation}
where $|0\rangle$ is the vacuum state. The energy functional, i.e.,
the energy density in nuclear matter, is then obtained by taking the
expectation of the Hamiltonian with respect to the ground state
$\lr\Phi_0\rc$ in a given volume $\Omega$,
 \begin{equation}\label{energy functional in nuclear
matter} \varepsilon=\frac{1}{\Omega}\left<\Phi_0\right|
H\left|\Phi_0\right>\equiv\left<{T}\right>+\sum_{\phi}\left<{V}_\phi\right>
= \varepsilon_k +\sum_{\phi}\lrb{
\varepsilon^D_\phi+\varepsilon^E_\phi },
\end{equation}
where $\phi=\sigma,\omega,\rho,\pi$, and
 \bsub\begin{eqnarray}
    \varepsilon_k &=& \sum_{p,s,\tau}\bar u(p,s,\tau)(\boldsymbol{\gamma}\cdot\boldsymbol{p}+M)u(p,s,\tau),\label{kinetic energy equation}\\
    \varepsilon^D_\phi &=& \frac{1}{2}\sum_{p_1,s_1,\tau_1}\sum_{p_2,s_2,\tau_2}
        \bar u(p_1,s_1,\tau_1)\bar u(p_2,s_2,\tau_2)\Gamma_\phi(1,2)\frac{1}{m^2_\phi}u(p_2,s_2,\tau_2)u(p_1,s_1,\tau_1),
        \label{direct energy equation}\\
\varepsilon^E_\phi &=&
-\frac{1}{2}\sum_{p_1,s_1,\tau_1}\sum_{p_2,s_2,\tau_2}
        \bar u(p_1,s_1,\tau_1)\bar
        u(p_2,s_2,\tau_2)\Gamma_\phi(1,2)\frac{1}{m^2_\phi+\boldsymbol{q}^2}u(p_1,s_1,\tau_1)u(p_2,s_2,\tau_2).
        \label{exchange energy equation}
\end{eqnarray}\esub

In the energy functional, $\varepsilon_k$ denotes the kinetic energy
density, and $\epsilon_\phi^D$ and $\varepsilon_\phi^E$ respectively
correspond to the direct (Hartree) and exchange (Fock) terms of the
potential energy density. With the Dirac spinors in
Eq.~(\ref{positive energy solutions}), one can obtain the
contributions of the energy density from each channel. The kinetic
energy density $\varepsilon_k$ and the direct terms of the potential
energy density $\varepsilon^D_\phi$ can be written as,
 \bsub\beqn\label{kinetic energy}
    \varepsilon_k &=&\sum_{i=n,p}\frac{1}{\pi^2}\int_0^{k_{F,i}} p^2 dp\lrb{ p\hat P+M\hat M}, \\
    \varepsilon^D_\sigma &=& -\frac{1}{2} \frac{g^2_\sigma}{m^2_\sigma}\rho^2_s,\\
    \varepsilon^D_\omega &=& +\frac{1}{2} \frac{g^2_\omega}{m^2_\omega}\rho^2_b,\\
    \varepsilon^D_\rho &=& +\frac{1}{2} \frac{g^2_\rho}{m^2_\rho}\rho_{b3}^2,
 \eeqn\esub
where the scalar density $\rho_s$,  baryonic density $\rho_b$ and the third component $\rho_{b3}$
read as,
 \bsub\begin{eqnarray}
    \rho_s &=&\sum_{i=n,p}\frac{1}{\pi^2}\int_0^{k_{F,i}}p^2dp\hat M(p),\\
    \rho_b &=&\sum_{i=n,p}\frac{k^3_{F,i}}{3\pi^2},\\
    \rho_{b3}&=&\frac{k^3_{F,n}}{3\pi^2}-\frac{k^3_{F,p}}{3\pi^2},
 \end{eqnarray}\esub
with the fermi momentum $k_{F,i}$ ($i = n,p$).

\begin{table}[htbp]\setlength{\tabcolsep}{6pt}
 \caption{The terms $A_\phi, B_\phi$ and
$C_\phi$ in Eq.~(\ref{eei})} \label{Tab:Eei}
 \begin{tabular}{c|c|c|c}
\hline\hline
 $\phi$ & $A_\phi(p,p')$ & $B_\phi(p,p')$ & $C_\phi(p,p')$\\
\hline
$\sigma$ & $g_\sigma^2\Theta_\sigma(p,p')$  & $g_\sigma^2\Theta_\sigma(p,p')$  & $-2g_\sigma^2\Phi_\sigma(p,p')$\\
$\omega$ & $2g_\omega^2\Theta_\omega(p,p')$ & $-4g_\omega^2\Theta_\omega(p,p')$& $-4 g_\omega^2\Phi_\omega(p,p')$\\
$\rho$   & $2g_\rho^2\Theta_\rho(p,p')$     & $-4g_\rho^2\Theta_\rho(p,p')$    & $-4  g_\rho^2\Phi_\rho(p,p')$\\
$\pi$    & $-f_\pi^2\Theta_\pi(p,p')$       & $-f_\pi^2\Theta_\pi(p,p')$       & $2\frac{f_\pi^2}{m_\pi^2}\left[\left(p^2+p'^2\right)\Phi_\pi(p,p')-pp'\Theta_\pi(p,p')\right]$\\
\hline\hline
 \end{tabular}
\end{table}

Compared to the simple form of direct terms of the potential energy
density, the exchange terms are much more complicated. In the
isoscalar channels ($\phi=\sigma,\omega$), the expressions read as,
\begin{eqnarray}\label{eei}
    \varepsilon^E_\phi=\frac{1}{2}\frac{1}{(2\pi)^4}\sum_{\tau,\tau'}\delta_{\tau\tau'}
        \int pp'dpdp'\lrs{A_\phi(p,p')+\hat M(p)\hat M(p')B_\phi(p,p')+\hat P(p)\hat P(p')C_\phi(p,p')}.
\end{eqnarray}
For the isovector channels ($\phi=\rho, \pi$), one just needs to
replace the isospin factor $\delta_{\tau\tau'}$ by
$\lrb{2-\delta_{\tau\tau'}}$ in the above expression. The details of
the terms $A_\phi,B_\phi$ and $C_\phi$ in Eq.~(\ref{eei}) are shown
in Table \ref{Tab:Eei}, where the functions $\Theta_\phi (p,p')$ and
$\Phi_\phi (p,p')$ are defined as,
 \bsub\begin{eqnarray}\label{integrals of propagators}
    \int d\Omega d\Omega'\frac{1}{m^2_\phi +\boldsymbol{q}^2}
    &=& \frac{4\pi^2}{pp'}\ln\frac{m^2_\phi +(p+p')^2}{m^2_\phi +(p-p')^2}
    \equiv\frac{4\pi^2}{pp'}\Theta_\phi (p,p'),\\
    \int d\Omega d\Omega'\frac{\hat{\svec p}\cdot\hat{\svec p}'}{m^2_\phi +\boldsymbol{q}^2}
    &=& \frac{4\pi^2}{pp'}\left\{\frac{p^2+p'^2+m^2_\phi }{2pp'}\Theta_\phi (p,p')-2\right\}
    \equiv2\frac{4\pi^2}{pp'}\Phi_\phi (p,p').
 \end{eqnarray}\esub

>From the potential energy densities in Eqs.~(\ref{direct energy
equation}) and (\ref{exchange energy equation}), one can perform
 the following variation,
\begin{equation}\label{self energy}
    \Sigma(p)u(p,s,\tau)=\frac{\delta}{\delta\bar u(p,s,\tau)}
        \sum_{\sigma,\omega,\rho,\pi}\left[ \varepsilon^D_\phi+\varepsilon^E_\phi\right],
\end{equation}
and obtain the self-energy $\Sigma(p)$ which includes the direct
terms,
\bsub\label{self energy of direct terms}\begin{eqnarray}
    \Sigma^D_S &=& -\frac{g_\sigma^2}{m_\sigma^2}\rho_s,\\
    \Sigma^D_0 &=& +\frac{g_\omega^2}{m_\omega^2}\rho_b+\frac{g_\rho^2}{m_\rho^2}\rho_{b3},
\end{eqnarray}\esub
and the exchange terms,
 \bsub
 \label{self energy of exchange terms}
\begin{eqnarray}
    \Sigma_{\tau,S}^E(p) &=&\frac{1}{(4\pi)^2 p}\int \hat M(p')p'dp' \sum_{\tau'}
    \left\{\delta_{\tau\tau'}\left[ B_\sigma + B_\omega\right]_{(p,p')}+
    (2-\delta_{\tau\tau'})\left[ B_\rho + B_\pi\right]_{(p,p')}\right\},\\
    \Sigma_{\tau,0}^E(p) &=&\frac{1}{(4\pi)^2 p}\int p'dp' \sum_{\tau'}
    \left\{\delta_{\tau\tau'}\left[ A_\sigma + A_\omega\right]_{(p,p')}+
    (2-\delta_{\tau\tau'})\left[ A_\rho + A_\pi\right]_{(p,p')}\right\},\\
    \Sigma_{\tau,V}^E(p) &=&\frac{1}{(4\pi)^2 p}\int \hat P(p')p'dp' \sum_{\tau'}
    \left\{\delta_{\tau\tau'}\left[ C_\sigma + C_\omega\right]_{(p,p')}+
    (2-\delta_{\tau\tau'})\left[ C_\rho + C_\pi\right]_{(p,p')}\right\}.
\end{eqnarray}
 \esub

In DDRHF, the explicit density dependence is introduced into the
meson-nucleon couplings, i.e., the coupling constants $g_\sigma$,
$g_\omega$, $g_\rho$ and $f_\pi$ are functions of the baryonic
density $\rho_b$. In the isoscalar meson-nucleon coupling channels

following form,
\begin{equation}
    g_\phi\left(\rho_b\right) = g_\phi\left(\rho_0\right)f_\phi \left(x\right),
\end{equation}
where $x=\rho_b/\rho_0$, and $\rho_0$ is the saturation density of nuclear matter, and the function
$f_\phi$ reads as
\begin{equation}
    f_\phi \left(x\right) = a_\phi \frac{1+b_\phi \left(x+d_\phi \right)^2}{1+c_\phi \left(x+d_\phi
    \right)^2}.
\end{equation}
In addition, five constraints, $f_\phi \left(1\right)=1$,
$f_\sigma''\left(1\right)=f_\omega''\left(1\right)$, and $f_\phi
''\left(0\right)=0$, are introduced to reduce the number of free
parameters. For the isovector channels, the exponential density
dependence is adopted for $g_\rho$ and $f_\pi$
\begin{eqnarray}
    g_\rho\left(\rho_b\right) &=& g_\rho\left(0\right)e^{-a_\rho x},\\
    f_\pi\left(\rho_b\right) &=& f_\pi\left(0\right)e^{-a_\pi x}.
\end{eqnarray}

Due to the density dependence in meson-nucleon couplings, the
additional contribution, i.e., the rearrangement term $\Sigma_R$,
appears in the self-energy $\Sigma$. In nuclear matter, it can be
written as,
\begin{equation}\label{SigS2}
    \Sigma_R=\sum_{\phi=\sigma,\omega,\rho,\pi}\frac{\partial g_\phi}{\partial \rho_b}
        \sum_\tau \frac{1}{\pi^2}\int \left[ \hat M(p)\Sigma^\phi_{\tau,S}(p)+\Sigma^\phi_{\tau,0}(p)
        +\hat P(p)\Sigma^\phi_{\tau,V}(p)\right] p^2dp.
\end{equation}

>From Eqs.~(\ref{self energy of direct terms}), (\ref{self energy of
exchange terms}) and (\ref{SigS2}) , the scalar component
$\Sigma_S$, time component $\Sigma_0$ and space component $\Sigma_V$
of the vector potential in Eq.~(\ref{Total Self Energy}) can be
obtained as,
 \bsub\begin{eqnarray}
    \Sigma_S(p) &=& \Sigma^D_S+\Sigma^E_S(p),\\
    \Sigma_0(p) &=& \Sigma^D_0+\Sigma^E_0(p)+\Sigma_R,\\
    \Sigma_V(p) &=& \Sigma^E_V(p),
\end{eqnarray}\esub
from which the starred quantities in Eq.~(\ref{Starred}) and the
hatted quantities in Eq.~(\ref{hatted quantities}) can be obtained.
Therefore, for nuclear matter with given baryonic density $\rho_b$
and neutron-proton ratio $N/Z$, one can proceed the self-consistent
iteration to investigate their properties: with the trial
self-energies, one can determine the starred quantities, hatted
quantities and calculate the scalar density, and then get the new
self-energies for next iteration until the final convergence.

In this work, a neutron star is described as the $\beta$-stable
nuclear matter system, which consists of not only neutrons and
protons, but also leptons $\lambda$ (mainly $e^-$ and $\mu^-$). The
equations of motion for the leptons are the free Dirac equations and
their densities can be expressed in terms of their corresponding
Fermi momenta, $\rho_\lambda=k_{F,\lambda}^3/(3\pi^2)$ ($\lambda =
e^-, \mu^-$). The chemical potentials of nucleons and leptons
satisfy the equilibrium conditions
 \begin{align}
\mu_p=&\mu_n-\mu_e, &\mu_{\mu}=&\mu_{e},
 \end{align}
where the chemical potentials $\mu_n$, $\mu_p$, $\mu_\mu$ and $\mu_e$ are determined by the
relativistic energy-momentum relation at the momentum $p = k_F$,
 \bsub\begin{align}
\mu_{i}=&\Sigma_0(k_{F,i})+E^*(k_{F,i}),\\
\mu_{\lambda}=&\sqrt{k_{F,\lambda}^2+m_\lambda^2},
 \end{align}\esub
where $ i = n,p$ and $\lambda = e^-, \mu^-$. The lepton masses are
respectively, $m_e=0.511$ MeV and $m_\mu=105.658$ MeV. In addition,
the baryon density conservation and charge neutrality are imposed as
 \begin{align}\label{densities&NPEM}
\rho_b=&\rho_n+\rho_p,&    \rho_p=&\rho_{\mu}+\rho_{e}.
 \end{align}

With these constraints, the energy density of neutron stars is then
obtained as
\begin{equation}
\varepsilon_{\rm ns}= \sum_{i=n,p,e,\mu}\varepsilon_{k,i} +\sum_{\phi=\sigma,\omega,\rho,\pi}\lrb{
\varepsilon^D_\phi+\varepsilon^E_\phi}.
\end{equation}
Here the leptons are treated as the free Fermi gas by assuming that there are no interactions
between leptons and nucleons or mesons and the kinetic energies of leptons can be expressed as,
 \begin{equation}
\varepsilon_{k,\lambda}=\frac{1}{\pi^2}\int_0^{k_{F,\lambda}}p^2dp\sqrt{p^2+m_\lambda^2}.
 \end{equation}
With the thermodynamic relation, the pressure of the neutron star system can be obtained as,
\begin{equation}
P({\rho_b}) = \rho_b^2 \frac{d}{d\rho_b}\frac{\varepsilon_{\rm
ns}}{\rho_b} = \sum_{i=n,p,e,\mu}\rho_i\mu_i-\varepsilon_{\rm ns}.
\end{equation}
At low density region  ($\rho_b<0.08$ fm$^{-3}$), instead of DDRHF calculations, the
BPS\cite{Baym:1971AJ} and BBP\cite{Baym:1971NPA} models are chosen to provide the proper EoS.

The structure equations of a static, spherically symmetric, relativistic star are the
Tolman-Oppenheimer-Volkov (TOV) equations \cite{Tolman:1939, Oppenheimer:1939}. Taking $c=G=1$, the
TOV equations read as
 \bsub\label{ToV}\begin{align}
    \frac{dP}{dr}=&-\frac{\left[P(r)+\varepsilon(r)\right]\left[M(r)+4\pi r^3 P(r)\right]}{r\left[r-2M(r)\right]},\\
    \frac{dM}{dr}=&4\pi r^2\varepsilon(r),
\end{align}\esub
where $P(r)$ is the pressure of the star at radius $r$, and $M(r)$
is the total star mass inside the sphere of radius $r$. Taking the
equation of state of stellar matter as the input, one could proceed
with the solution of TOV equations. The point $R$, at which the
pressure vanishes, i.e., $P(R)=0$, defines the radius of the star
and the corresponding $M(R)$ is the gravitational mass. For a given
EoS, the TOV equation has the unique solution which depends on a
single parameter characterizing the conditions of matter at the
center, such as the central density $\rho(0)$ or the central
pressure $P(0)$.

\section{Results and discussions}\label{sec:RD}
In this paper, the EoS and the neutron star properties are studied
in DDRHF with the effective interactions PKO1, PKO2 and PKO3
\cite{Long:2006, Long:2008}. As shown in Table \ref{Tab:RHFS}, the
coupling constant $g_\rho(0)$ is fixed to the value in the free
space in PKO1 whereas free to be adjusted in PKO2 and PKO3, and
$\pi$-coupling is not included in PKO2. For comparison, the results
calculated by RMF are also discussed. The effective interactions
used in RMF calculations include the nonlinear self-coupling ones
GL-97 \cite{gle}, NL1 \cite{Reinhard:1986}, NL3
\cite{Lalazissis:1997}, NLSH \cite{Sharma:1993}, TM1
\cite{Sugahara:1994} and PK1 \cite{Long04}, and the
density-dependent ones TW99 \cite{Typel:1999}, DD-ME1
\cite{Niksic:2002}, DD-ME2 \cite{Lalazissis:2005} and PKDD
\cite{Long04}.

\begin{table}[htbp]\setlength{\tabcolsep}{6pt}
 \caption{The effective interactions PKO1, PKO2 and PKO3 of DDRHF\cite{Long:2006}, where the masses $M =
938.9$ MeV, $m_\omega = 783.0$ MeV, $m_\rho = 769.0$ MeV and $m_\pi
= 138.0$ MeV.} \label{Tab:RHFS}
 \begin{tabular}{ccccccccc}
\hline\hline
&$m_\sigma$&$g_\sigma$&$g_\omega$&$g_\rho(0)$&$f_\pi(0)$&$a_\rho$&$a_\pi$&$\rho_0$\\
\hline
PKO1&525.769084&8.833239&10.729933&2.629000&1.000000&0.076760&1.231976&0.151989\\
PKO2&534.461766&8.920597&10.550553&4.068299&$-$&0.631605&$-$&0.151021\\
PKO3&525.667686&8.895635&10.802690&3.832480&1.000000&0.635336&0.934122&0.153006\\
\hline\hline
&$a_\sigma$&$b_\sigma$&$c_\sigma$&$d_\sigma$&$a_\omega$&$b_\omega$&$c_\omega$&$d_\omega$\\
\hline
PKO1&1.384494&1.513190&2.296615&0.380974&1.403347&2.008719&3.046686&0.330770\\
PKO2&1.375772&2.064391&3.052417&0.330459&1.451420&3.574373&5.478373&0.246668\\
PKO3&1.244635&1.566659&2.074581&0.400843&1.245714&1.645754&2.177077&0.391293\\
\hline\hline
 \end{tabular}
\end{table}

\subsection{Properties of nuclear matter}\label{sec:RDNM}

\subsubsection{Bulk properties}
In Table \ref{Tab:Sat} are shown the bulk quantities of nuclear
matter at saturation point, i.e., the saturation density $\rho_0$,
the binding energy per particle $E_B/A$, the incompressibility $K$,
the symmetry energy $J$ and the scalar mass $M_S^*/M$. The results
calculated by RMF with both the nonlinear self-coupling effective
interactions and the density-dependent ones, which have been studied
systematically in Ref. \cite{Ban:2004}, are included for comparison.
The saturation density and the binding energy per particle given by
DDRHF with PKO series are around 0.152 fm$^{-3}$ and $-16.0$ MeV,
respectively, close to the values provided by RMF. The
incompressibility $K$ calculated by DDRHF with PKO1, PKO2 and PKO3
range from 249 to 262 MeV, close to the values given by RMF with
density dependent effective interaction. In contrast, relatively
large values of $K$ (270$\sim$360 MeV) are obtained by RMF with the
non-linear self-coupling of mesons, except GL-97 and NL1. For the
symmetry energy $J$, the non-linear version of RMF also presents
relatively large values (36$\sim$44 MeV) except GL-97, whereas the
density-dependent version of RMF (except PKDD) provides comparative
values (32$\sim$34 MeV) to DDRHF with PKO1, PKO2 and PKO3. For the
scalar mass $M_S^*$, GL-97 gives the largest value and TW99 presents
the smallest. The values given by DDRHF with PKO series are around
0.60, close to those by RMF with the non-linear self-couplings of
mesons except GL-97, and systematically smaller values are obtained
by RMF with the density dependent meson-nucleon couplings.

 \begin{table}[htbp]\setlength{\tabcolsep}{8pt}
\caption{The saturation density $\rho_0$ (fm$^{-3}$), binding energy per particle $E_B/A$ (MeV),
incompressibility $K$ (MeV), asymmetry energy coefficient $J$ (MeV) and the scalar mass $M_S^*/M$
of nuclear matter at saturation point. } \label{Tab:Sat}
 \begin{tabular}{c*{5}{c}}\hline\hline
       &$\rho_0$ & $E_B/A$ & $K$     & $J$    & $M^*_S/M$\\ \hline
 PKO1  & 0.1520  & -15.996 & 250.239 & 34.371 & 0.5900   \\
 PKO2  & 0.1510  & -16.027 & 249.597 & 32.492 & 0.6025   \\
 PKO3  & 0.1530  & -16.041 & 262.469 & 32.987 & 0.5862   \\ \hline\hline
 GL-97 & 0.1531  & -16.316 & 240.050 & 32.500 & 0.7802   \\
 NL1   & 0.1518  & -16.426 & 211.153 & 43.467 & 0.5728   \\
 NL3   & 0.1483  & -16.249 & 271.730 & 37.416 & 0.5950   \\
 NLSH  & 0.1459  & -16.328 & 354.924 & 36.100 & 0.5973   \\
 TM1   & 0.1452  & -16.263 & 281.162 & 36.892 & 0.6344   \\
 PK1   & 0.1482  & -16.268 & 282.694 & 37.642 & 0.6055   \\ \hline
 TW99  & 0.1530  & -16.247 & 240.276 & 32.767 & 0.5549   \\
 DD-ME1& 0.1520  & -16.201 & 244.719 & 33.065 & 0.5780   \\
 DD-ME2& 0.1518  & -16.105 & 250.296 & 32.271 & 0.5722   \\
 PKDD  & 0.1496  & -16.268 & 262.192 & 36.790 & 0.5712   \\ \hline\hline
 \end{tabular}
 \end{table}

\subsubsection{Density dependence of the coupling constants}

\begin{figure}[htbp]
\includegraphics[width=0.80\textwidth]{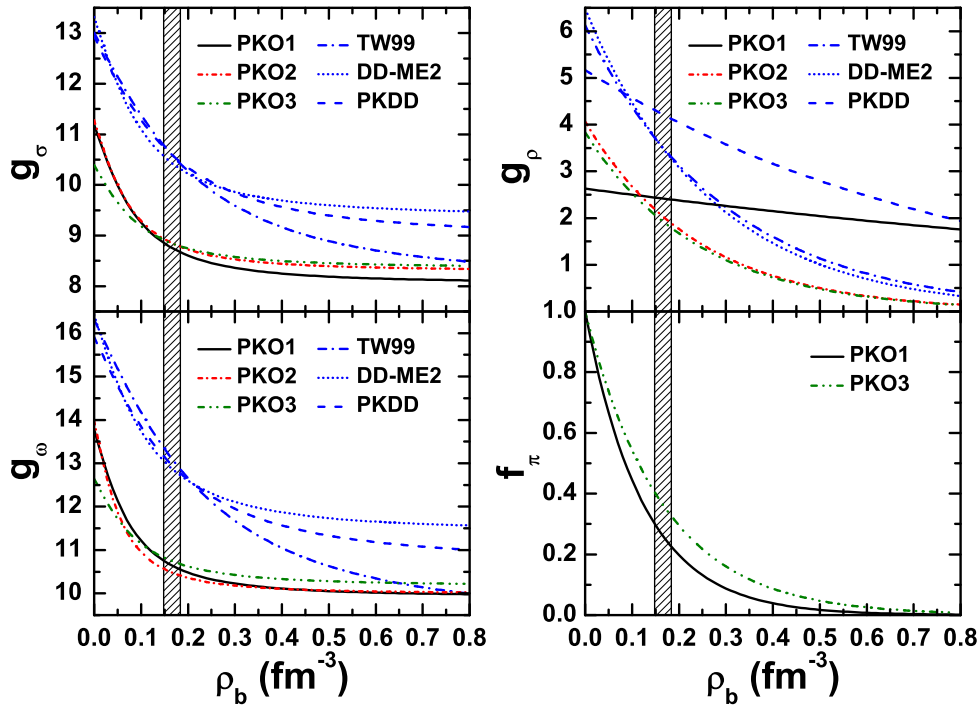}
\caption{(Color online) The coupling constants $g_\sigma$,
$g_\omega$, $g_\rho$ and $f_\pi$ as functions of the baryonic
density $\rho_b$ (fm$^{-3}$) for the DDRHF effective interactions
PKO1, PKO2 and PKO3, and RMF ones PKDD, TW99 and DD-ME2. The
shadowed area represents the empirical saturation region
$\rho_b=0.166\pm0.018~{\rm fm^{-3}}$.} \label{Fig:Constant}
 \end{figure}

In DDRHF, the medium effects are evaluated by the density dependence in the meson-nucleon
couplings. To understand the EoS, it is worthwhile to have a look at the density dependence of the
coupling constants. In Fig. \ref{Fig:Constant} are shown the coupling constants $g_\sigma$,
$g_\omega$, $g_\rho$ and $f_\pi$ as functions of baryonic density $\rho_b$, where the results of
the DDRHF effective interactions PKO1, PKO2 and PKO3 are given as compared to the RMF ones TW99,
DD-ME2 and PKDD. As seen from Fig. \ref{Fig:Constant}, all the effective interactions present
strong density dependence in the low density region ($\rho_b<0.2$ fm$^{-3}$) for both isoscalar
($\sigma$ and $\omega$) and isovector ($\rho$ and $\pi$) meson-nucleon couplings. When density
grows higher, $g_\sigma$ and $g_\omega$  in the left panels become stable. While due to the
exponential density dependence, the isovector ones $g_\rho$ and $f_\pi$ tend to vanish (except
$g_\rho$ in PKDD and PKO1) as shown in the right panels. From this aspect, one can understand that
the isoscalar mesons provide the dominant contributions in the high density region. Compared to the
RMF effective interactions, PKO1, PKO2 and PKO3 have smaller $g_\sigma$ and $g_\omega$. This is
mainly due to the effects of Fock terms, which lead to the recombination of the ingredients in the
nuclear interactions. With the inclusion of Fock terms, the nuclear attractions are shared by the
Hartree terms of $\sigma$-coupling and the Fock terms of $\omega$-, $\rho$- and $\pi$-couplings,
and the repulsions are contributed by the Hartree terms of $\omega$-coupling and the Fock terms of
$\sigma$-coupling. While in RMF, the attraction and repulsion are provided only by the Hartree
terms of the $\sigma$- and $\omega$-couplings, respectively.

It is not enough to adjust the isospin properties only within the
nuclear saturation region. It is expected that the investigations on
the EoS at higher densities and neutron star properties could
provide the additional constraint. For the isovector coupling
constants in the right panels of Fig.~\ref{Fig:Constant}, PKO1 and
PKDD present slightly weak density dependence in $g_\rho$ because of
the fairly small density dependent parameter $a_\rho$. In analogy to
$g_\sigma$ and $g_\omega$, the RMF effective interactions give
larger values of $g_\rho$. This is also due to the exchange
contributions. In DDRHF, significant contributions to the isospin
part of nuclear interaction are found in exchange terms of isovector
mesons as well as isoscalar ones. It is different from the situation
in RMF that the isospin properties are only described by the direct
part of $\rho$-coupling. For the $\pi$-meson, the contribution in
neutron stars is negligible since $f_\pi$ tends to vanish at high
densities.

\subsubsection{Equations of state}

\begin{figure}[htbp]
 \includegraphics[width=0.48\textwidth]{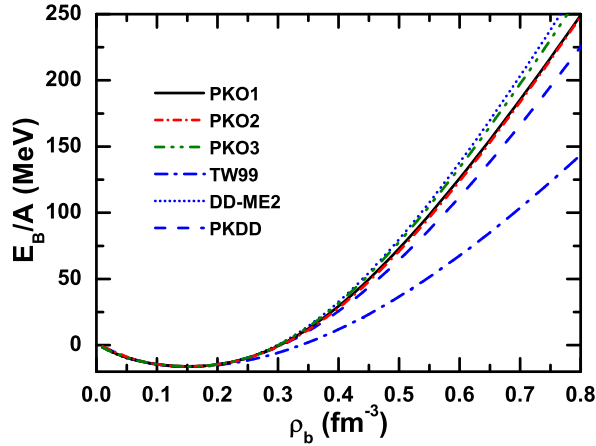}
\caption{(Color online) The binding energy per particle $E_B/A$ as
function of the baryonic density $\rho_b$ for symmetric nuclear
matter. The results are calculated by DDRHF with PKO1, PKO2 and
PKO3, in comparison with those by RMF with TW99, DD-ME2 and PKDD. }
\label{Fig:EbSNM}
\end{figure}

\begin{figure}[htbp]
 \includegraphics[width=0.48\textwidth]{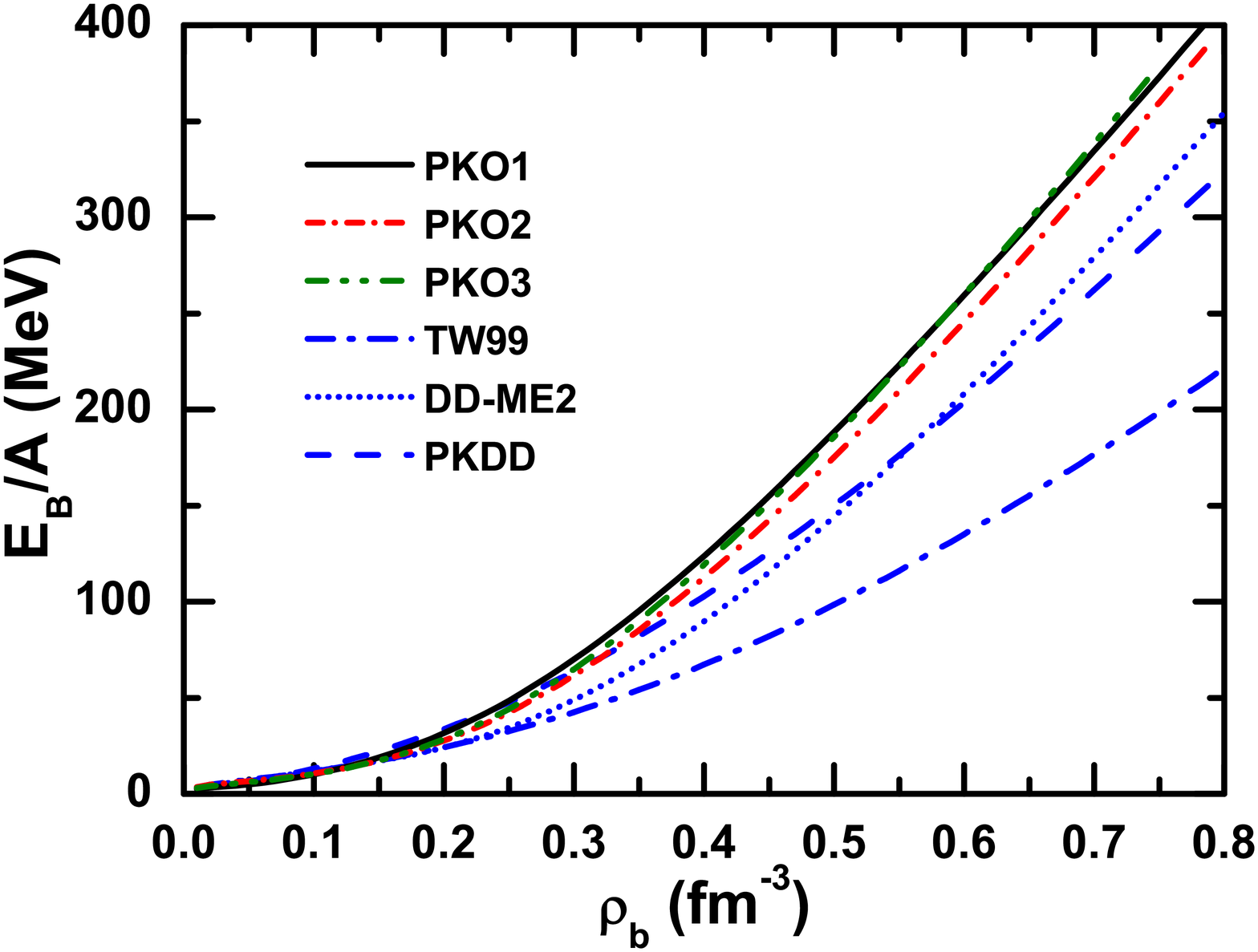}
\caption{(Color online) Similar to Fig. \ref{Fig:EbSNM} but for pure
neutron matter.} \label{Fig:EbPNM}
\end{figure}

The equations of state calculated by DDRHF with PKO1, PKO2 and PKO3
are shown in Fig. \ref{Fig:EbSNM} and Fig. \ref{Fig:EbPNM},
respectively for the symmetric nuclear matter and pure neutron
matter. The results calculated by RMF with TW99, DD-ME2 and PKDD are
also shown for comparison. It is recommended to see Ref.
\cite{Ban:2004} for the density dependence of the EoS on more RMF
effective interactions. Seen from these two figures, identical
behaviors of the EoS are provided by all the effective interactions
at low density region ($\rho_b<\rho_0$) while in high density region
exist pronounced deviations among different effective interactions.

For the symmetric nuclear matter in Fig. \ref{Fig:EbSNM}, DDRHF with PKO1, PKO2 and PKO3 provides
similar EoS as RMF with PKDD and DD-ME2, whereas much softer EoS is obtained by RMF with TW99 when
density grows high. For the pure neutron matter in Fig. \ref{Fig:EbPNM}, the curves can be
classified into three groups according to the behaviors of the EoS at high density region. Among
all the effective interactions, the DDRHF ones present the hardest equations of state and the RMF
one TW99 gives the softest one, whereas DD-ME2 and PKDD provide similar equations of states, which
lie between the hardest and softest. Since the DDRHF parameterizations were performed by fitting
the properties of finite nuclei and nuclear matter around saturation point \cite{Long:2006}, which
corresponds to the low density region, it becomes necessary to test the extrapolation of the
effective interactions PKO1, PKO2 and PKO3 to high densities.

\subsubsection{Symmetry energy}

The EoS property of isospin asymmetric nuclear matter is still
ambiguous more or less. Different theoretical models predict quite
different behaviors of the EoS for pure neutron matter. In most
cases, it is due to the effective interactions obtained by fitting
the properties of doubly magic nuclei, which have an isospin close
to that of the symmetric nuclear matter. From this point of view, it
becomes necessary to introduce the constraints, either from isospin
asymmetric heavy-ion collisions experiments or from the data of
nuclei with extreme isospin, into the fitting procedures of the
effective interactions.

The symmetry energy is an important quantity to illustrate the
property of asymmetric nuclear matter. In general, the energy per
particle of asymmetric nuclear matter $E(\rho_b,\beta)$ can be
expanded in a Taylor series with respect to $\beta$,
\begin{equation}\label{EsParabolicLaw}
E(\rho_b,\beta)=E_0(\rho_b)+\beta^2E_S(\rho_b)+\cdots,
\end{equation}
where $\beta=1-2\rho_p/\rho_b$ is the asymmetry parameter depending
on the proton fraction. The function $E_0(\rho_b)$ is the binding
energy per particle in symmetric nuclear matter, and the symmetry
energy $E_S(\rho_b)$ ($J = E_S(\rho_0)$) is denoted as
 \begin{equation}
E_S(\rho_b)=\left.\frac{1}{2}\frac{\partial^2E(\rho_b,\beta)}{\partial\beta^2}\right|_{\beta=0}.
 \end{equation}
The empirical parabolic law in Eq.~(\ref{EsParabolicLaw}) is
confirmed to be reasonable in all the range of the asymmetry
parameter values, while at high density deviation from such a
behavior is found \cite{Bombaci:1991}.

\begin{figure}[htbp]
 \includegraphics[width=0.48\textwidth]{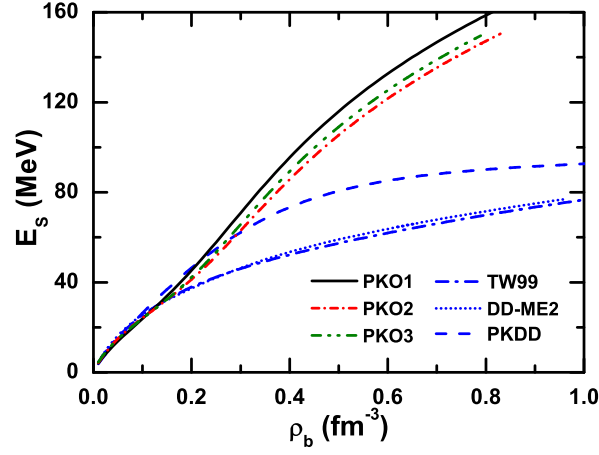}
\caption{(Color online) The nuclear symmetry energy $E_S$ (MeV) as a
function of the baryon density $\rho_b$ (fm$^{-3}$). The results are
calculated by DDRHF with PKO1, PKO2 and PKO3, in comparison to those
by RMF with TW99, DD-ME2 and PKDD.} \label{Fig:Asym}
\end{figure}

Fig. \ref{Fig:Asym} shows the symmetry energy as a function of the
baryon density $\rho_b$. The results are calculated by DDRHF with
PKO1, PKO2 and PKO3, in comparison with those by RMF with TW99,
DD-ME2 and PKDD. As shown in Fig. \ref{Fig:Asym}, both DDRHF and RMF
effective interactions present identical behaviors of the symmetry
energy at low densities ($\rho<\rho_0$), while sizeable enhancements
in high density region are obtained by DDRHF with PKO1, PKO2 and
PKO3 as compared to the RMF results. Among the RMF calculations,
PKDD shows harder behavior than DD-ME2 and TW99, which provide
identical symmetry energy in the whole density region.

\begin{figure}[htbp]
\includegraphics[width = 0.45\textwidth]{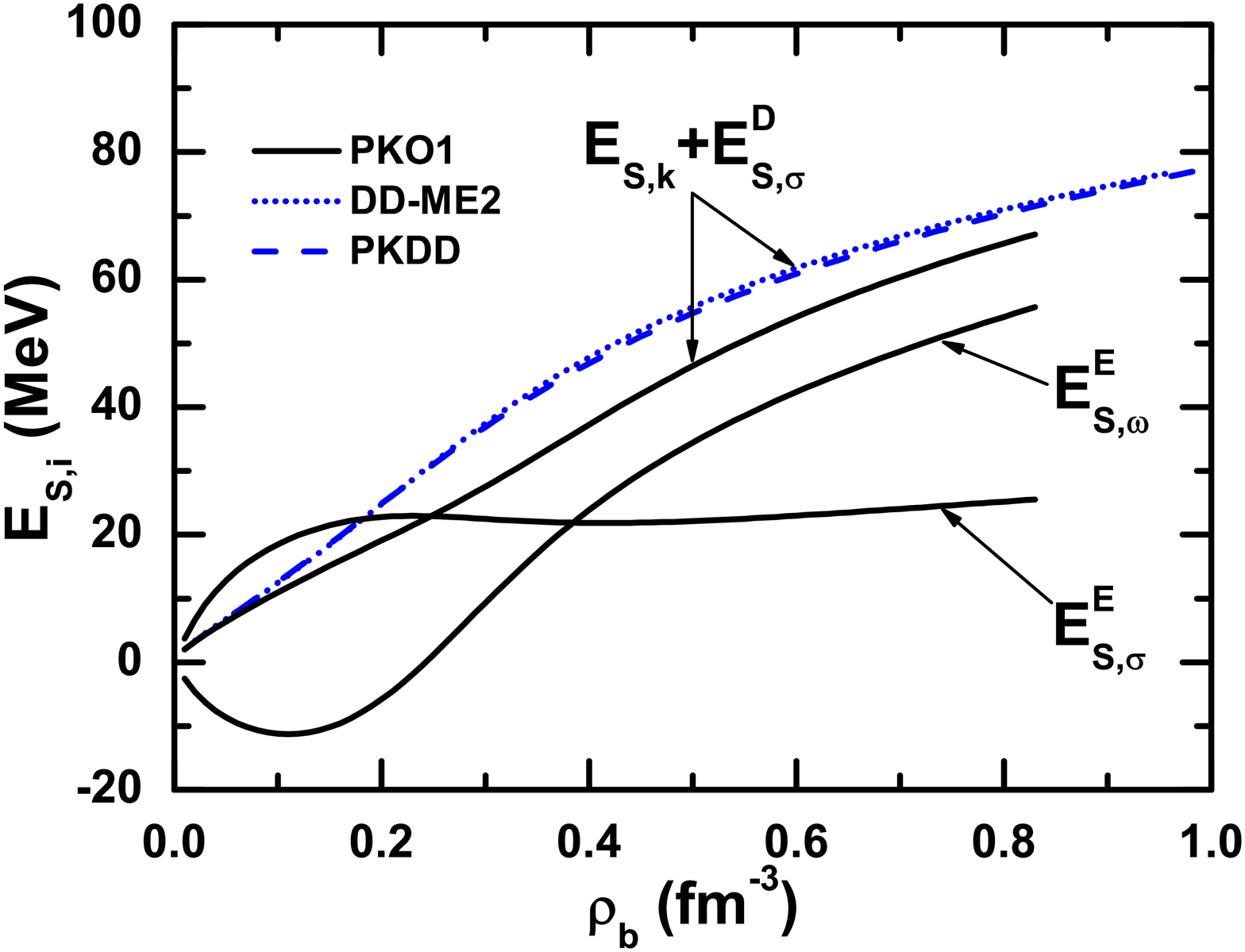}~~~
\includegraphics[width = 0.45\textwidth]{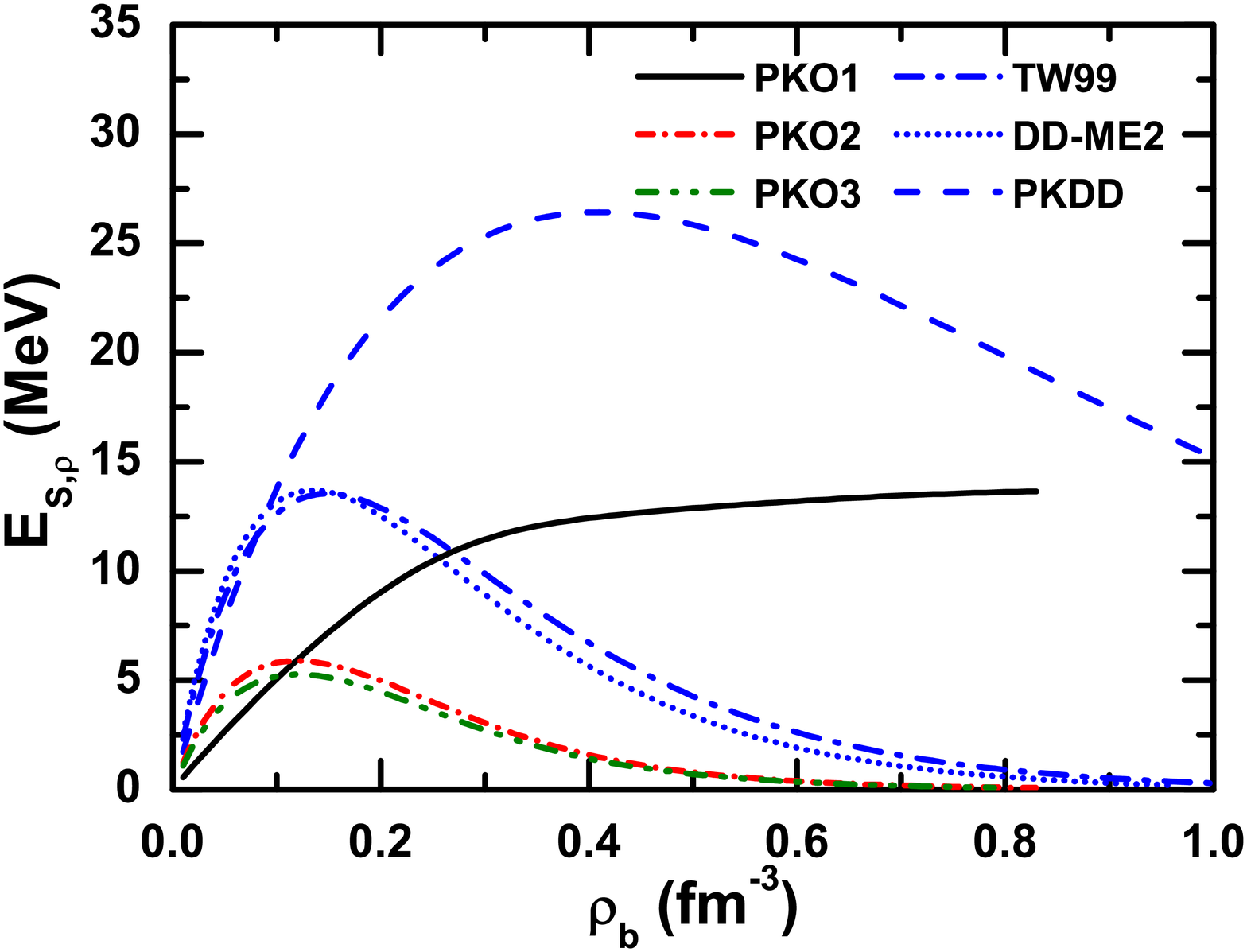}
\caption{(Color online) Contributions from different channels to the
symmetry energy as a function of the baryon density $\rho_b$. Left
panel gives the contributions from the kinetic energy and isoscalar
channels, and the ones from $\rho$-meson are shown in the right
panel. See text for details. }
 \label{Fig:Apart&rho}
\end{figure}

>From the energy functional in nuclear matter in Eq.~(\ref{energy
functional in nuclear matter}), one can obtain the contributions
from different channels to the symmetry energy $E_S$ as,
 \begin{equation}
E_S=E_{S,k}+ \sum_\phi \lrb{E_{S,\phi}^D + E_{S,\phi}^E},
 \end{equation}
where $\phi = \sigma, \omega, \rho$ and $\pi$. In fact, the direct
terms of $\omega$-meson coupling have no contribution to the
symmetry energy because of the nature of isoscalar-vector coupling.
It is also expected that the one-pion exchange has minor effects
since nuclear matter is a spin-saturated system. In Fig.
\ref{Fig:Apart&rho}, the contributions from different channels to
the symmetry energy are shown as functions of the baryon density
$\rho_b$. In the left panel are presented the contributions from the
kinetic part and isoscalar channels, and only the results calculated
by DDRHF with PKO1 are shown in comparison with those by RMF with
PKDD and DD-ME2. The contributions from the $\rho$-meson coupling
are shown in the right panel, including the results calculated by
DDRHF with PKO1, PKO2 and PKO3, and RMF with PKDD, DD-ME2 and TW99.

Within RMF, one can find that the kinetic part and the direct terms
of $\sigma$ coupling, i.e., $E_{S,k}$ and $E_{S,\sigma}^D$, provide
the dominant contributions to the symmetry energy. As shown in the
left panel of Fig. \ref{Fig:Apart&rho}, PKDD and DD-ME2 give
identical values of $E_{S,k}+E_{S,\sigma}^{D}$, and the deviation
between them appearing in Fig. \ref{Fig:Asym} is mainly due to their
contributions from the $\rho$-coupling as seen from the right panel
of Fig. \ref{Fig:Apart&rho}.  In the left panel of Fig.
\ref{Fig:Apart&rho}, the values of $E_{S,k}+E_{S,\sigma}^{D}$ given
by the DDRHF calculations are found smaller than those of RMF with
PKDD and DD-ME2. It is also seen that the Fock terms of $\sigma$-
and $\omega$-couplings present significant contributions to the
symmetry energy, which well interprets the stronger density
dependence predicted by DDRHF than by RMF at the high density region
(see Fig. \ref{Fig:Asym}). As seen from the left panel of Fig.
\ref{Fig:Apart&rho}, the values of $E_{S,\sigma}^E$ increase rapidly
in the low density region and tend to be stable about $20\sim25$MeV
at high density. While the exchange terms of $\omega$-coupling
provide negative contributions to the symmetry energy at low density
and reach the minimum about $-11$ MeV at $\rho_b = 0.1$ fm$^{-3}$.
After that the values of $E_{S,\omega}^E$ grow up and become
comparable to the values of $E_{S,k}+E_{S,\sigma}^{D}$ at several
times of the saturation density.

In the right panel of Fig. \ref{Fig:Apart&rho}, the contributions of
$\rho$-coupling, i.e., $E_{S,\rho} = E_{S,\rho}^D + E_{S,\rho}^E$,
are found to be important for the symmetry energy at low density
region. When density goes high, the values of $E_{S,\rho}$ given by
all the effective interactions except PKO1 and PKDD tend to zero due
to their strong exponential density dependence in $\rho$-nucleon
coupling. Because of much smaller value of $a_\rho$, PKO1 presents
larger contributions than PKO2 and PKO3 and contributes a value
about $10\sim15$ MeV in the high density region. Due to the same
reason, PKDD also provides larger values of $E_{S,\rho}$ than other
two RMF effective interactions, and it reaches the maximum about 26
MeV at $\rho_b\simeq0.41$ fm$^{-3}$, then falls down slowly.
Comparing the values of $E_{S,\rho}$ given by PKO2 and PKO3 to those
by DD-ME2 and TW99, the contribution of $\rho$-coupling $E_{S,\rho}$
is depressed systematically in DDRHF. Such kind of depressions also
exist between the results of PKO1 and PKDD, which are of similar
density dependence in $\rho$-nucleon coupling. This could be
understood from the fact that smaller values of $g_\rho$ are
obtained with the inclusion of Fock terms as seen in Fig.
\ref{Fig:Constant}. In fact, not only the $\rho$-meson but all the
mesons take part to the isospin properties and are in charge of
producing the symmetry energy via the Fock channel.

Concluding the above discussions, one can find that the Fock terms
play an important role in determining the density dependent behavior
of the symmetry energy. It is then expected that the important
constraints on the symmetry energy and the EoS of asymmetric nuclear
matter could be obtained from the study of neutron stars.

\subsection{Properties of neutron star}\label{sec:RDNS}
In this work, the static and $\beta$-equilibrium assumptions are
imposed for the description of neutron stars. With the density
increasing, the high momentum neutrons will $\beta$ decay into
protons and electrons, i.e., $n\leftrightarrow p+e^-+\bar{\nu_e}$,
until the chemical potentials satisfy the equilibrium
$\mu_p=\mu_n-\mu_e$. When the chemical potential of electron $\mu_e$
reaches the limit of the muon mass, the lepton $\mu^-$ will appear.
The reaction $e^- \leftrightarrow \mu^-+\bar{\nu_\mu}+\nu_e$ implies
the equilibrium between the $e^-$ and $\mu^-$ chemical potentials,
i.e., $\mu_e=\mu_\mu$.

\subsubsection{Density distribution}

\begin{figure}[hbtp]
 \includegraphics[width=0.75\textwidth]{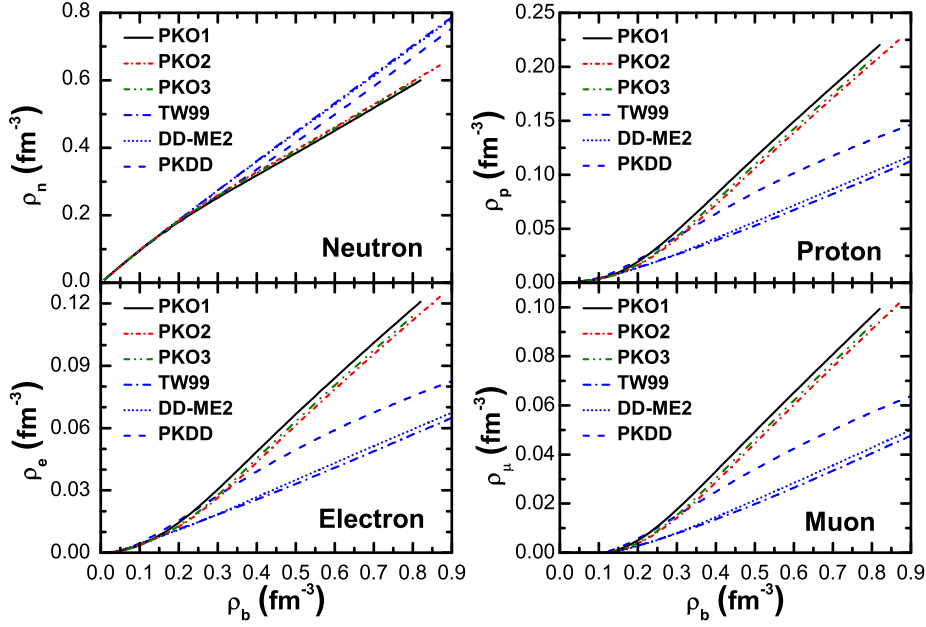}
\caption{(Color online) The neutron (up-left panel), proton
(up-right panel), electron (lower-left panel) and muon (lower-right
panel) densities in neutron star matter as functions of the baryon
density $\rho_b$ (fm$^{-3}$). The results are calculated by DDRHF
with PKO1, PKO2 and PKO3, in comparison to those by RMF with TW99,
DD-ME2 and PKDD.} \label{Fig:Density}
\end{figure}

To keep the equilibrium among the particle chemical potentials,
protons, electrons and muons will appear with the density increasing
in neutron stars. In Fig.~\ref{Fig:Density} are shown the neutron,
proton, electron and muon densities in neutron stars as functions of
the baryon density. The results are calculated by DDRHF with PKO1,
PKO2 and PKO3, in comparison to those by RMF with TW99, DD-ME2 and
PKDD. The density distribution of various components in RMF with
both the nonlinear self-coupling effective interactions and the
density-dependent ones have been studied systematically in Ref.
\cite{Ban:2004}. As seen from Fig. \ref{Fig:Density}, the thresholds
of $\mu^-$ occurrence predicted by different effective interactions
are very close to each another, roughly around $\rho_b=0.12~{\rm
fm}^{-3}$. It is shown that all the densities keep increasing
monotonously with respect to the baryonic density $\rho_b$. Similar
as the situation in the EoS of nuclear matter, different effective
interactions present identical trends at low densities
($\rho_b<\rho_0$), while remarkable deviations exist in high density
region between the DDRHF and RMF predictions. Seen from Fig.
\ref{Fig:Density}, the results given by different effective
interactions can be classified into three groups, the DDRHF ones,
PKDD, and the RMF ones DD-ME2 and TW99. For the proton, electron and
muon densities, the strongest density dependence is presented by
DDRHF with PKO1, PKO2 and PKO3, while the softest behaviors are
provided by RMF with DD-ME2 and TW99. In contrast, the softest
behavior on neutron density is predicted by the DDRHF effective
interactions, whereas TW99 and DD-ME2 present the hardest. This kind
of reversion can be well understood from the relations in
Eq.~(\ref{densities&NPEM}) among the densities.

It is known that the density fractions of each components in neutron
stars are rather sensitive to the symmetry energy, as illustrated by
associating Fig.~\ref{Fig:Density} with Fig.~\ref{Fig:Asym}. Due to
the strong effects from the exchange terms of $\omega$-coupling in
high density region (see Fig.~\ref{Fig:Apart&rho}), DDRHF with PKO1,
PKO2 and PKO3 shows stronger density dependence on the symmetry
energy and then are obtained the harder behaviors on the proton,
electron and muon density distributions, as compared to the RMF
calculations. For the deviations between different effective
interactions within one theoretical model, e.g., between PKDD and
DD-ME2, they are mainly due to the $\rho$-coupling as shown in the
right panel of Fig. \ref{Fig:Apart&rho}, where the $\rho$-meson
coupling of PKDD shows larger contributions to the symmetry energy.
As a conclusion, the harder behavior on the symmetry energy at high
densities, more difficult the system becomes asymmetric and more
easier neutrons decay into protons and electrons, which leads to
smaller neutron abundance and larger proton, electron and muon
abundances in neutron stars.

\subsubsection{Proton fraction and direct Urca constraint}
\begin{figure}[hbtp]
 \includegraphics[width=0.48\textwidth]{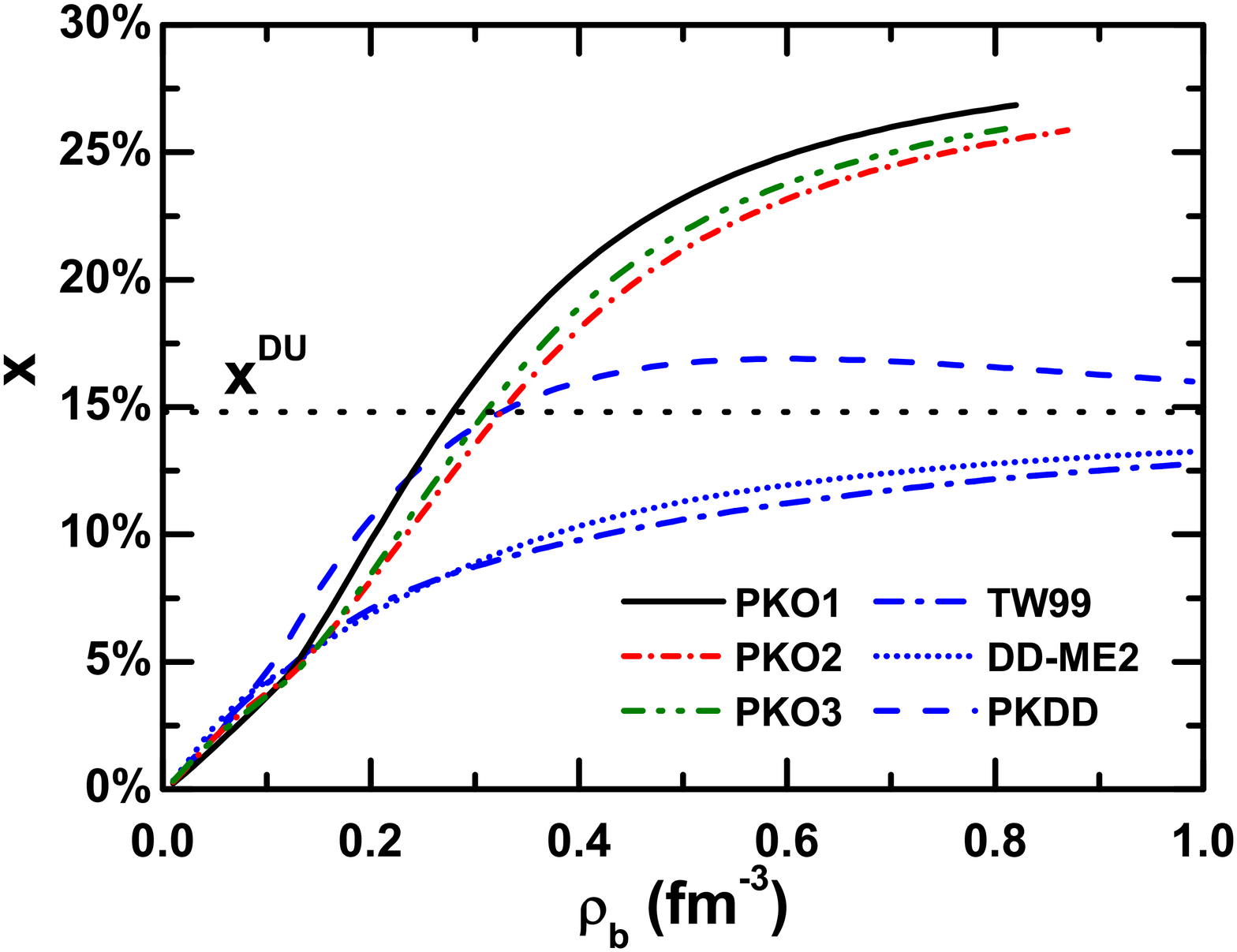}
 \caption{(Color online) Proton fractions $x=\rho_p/(\rho_p+\rho_n)$ in neutron star
matter for different DDRHF and DDRMF effective interactions. The
dotted line labeled with $x^{DU}$ is threshold for happening direct
Urca process. Here $x^{DU}=14.8\%$ is taken by assuming muons in the
massless limit.}
 \label{Fig:Fraction}
\end{figure}

>From the density distributions in Fig. \ref{Fig:Density}, one can
extract the proton fraction $x=\rho_p/(\rho_p+\rho_n)$ within the
range of density of neutron stars. Fig. \ref{Fig:Fraction} shows the
proton fraction $x$ as a function of baryonic density $\rho_b$,
where the results calculated by DDRHF with PKO series are presented
in comparison to those by RMF with TW99, DD-ME2 and PKDD. Due to the
stiff behavior on the symmetry energy (see Fig. \ref{Fig:Asym}),
stronger density dependence of the proton fraction $x$ in neutron
star matter is obtained by DDRHF than RMF as shown in Fig.
\ref{Fig:Fraction}.

The cooling mechanism of neutron stars, which is sensitive to the
proton fraction, could bring significant information of asymmetric
nuclear EoS. Direct Urca (DU) processes $n\rightarrow
p+e^-+\bar{\nu}_e$ and $p+e^-\rightarrow n+\nu_e$ lead the star to
cool off rapidly by emitting the thermal neutrinos. The threshold of
the proton fraction $x^{DU}$ for the DU process occurring can be
easily found as $11.1\%\leqslant x^{DU}\leqslant 14.8\%$ with the
momentum conservation and charge neutrality \cite{Lattimer:1991,
Klahn:2006ir}. Seen From Fig. \ref{Fig:Fraction}, the critical
density $\rho^{DU}$ for the DU process occurring depends on the EoS.
Once the critical density $\rho^{DU}$ is reached in the center of a
neutron star for a given EoS, the star will be efficiently cooled
via the DU process. It is found that the values of $x^{DU}$ given by
DDRHF calculations correspond to fairly low critical densities while
the results calculated by RMF with TW99 and DD-ME2 do not support
the DU process occurring at all. The DU critical star masses
$M^{DU}$ and corresponding central densities $\rho^{\rm DU}(0)$ are
marked in Fig. \ref{Fig:Mass} by filled squares.

According to the analysis in Refs. \cite{Blaschke:2004, Popov:2006,
Klahn:2006ir}, if the DU process is taken as a possible mechanism
for neutron star cooling, an acceptable EoS shall not allow it to
occur in neutron stars with masses below 1.5 $M_\odot$, otherwise it
will be in disagreement with modern observational soft X-ray data in
the temperature-age diagram. As a weaker constraint, the limit
$M^{\rm DU}>1.35~M_\odot$ could be applied. From the mass limit
$M^{DU}$, are then obtained the constraint over the EoS that the
density dependence of the symmetry energy should not be too strong,
and probably not too weak, either. In Table \ref{Tab:cooling}, are
given the critical neutron star mass $M^{DU}$ and central densities
$\rho^{\rm DU}(0)$ from the DDRHF and RMF calculations, which
support the occurrence of the DU cooling process in stars.

Seen from Table \ref{Tab:cooling}, rather small mass limits $M^{DU}$
are obtained by RMF with the non-linear self-coupling of mesons
while the DDRHF calculations with PKO2 and PKO3 provide larger
values of $M^{DU}$, which are very close to the limit 1.5 $M_\odot$
mentioned above and satisfy the weak constraint that $M^{\rm
DU}>1.35~M_\odot$. For the calculation with PKO1, the DU cooling
process will occur at the fairly low mass 1.20 $M_\odot$ and central
density $\rho^{DU}\simeq0.28$ fm$^{-3}$, which can be interpreted by
the contributions of the $\rho$-meson coupling to the symmetry
energy. For the $E_{S,\rho}$ in the right panel of Fig.
\ref{Fig:Apart&rho}, the $\rho$-meson coupling in PKO1 still has
remarkable effects in the high density region due to the weak
density dependence of $g_\rho$ (see Fig. \ref{Fig:Constant}). Due to
the same reason, the RMF calculation with PKDD also support the DU
cooling process to occur at a low mass limit 1.26 $M_\odot$. In
contrast, as seen in Fig. \ref{Fig:Fraction}, the occurrence of the
DU cooling process is not supported at all by the RMF calculations
with TW99 and DD-ME2 as well as DD-ME1. It is expected that the
occurrence of the DU process could be introduced as a possible
constraint in the future parameterizations of both DDRHF and RMF,
e.g., for the $\rho$-meson coupling ($g_\rho(0)$ and $a_\rho$).

\begin{table}[htbp] \setlength{\tabcolsep}{0.5em}
\caption{Critical neutron star masses $M^{\rm DU}$ and central
densities $\rho^{\rm DU}(0)$ for the occurrence of the DU cooling
process and the criterion of the DU constraint given by both DDRHF
and RMF effective interactions. Fulfillment (violation) of a
constraint is indicated with $+(-)$.}
 \begin{tabular}{c|*{3}{c}|*{6}{c}|c}
\hline\hline
 & PKO1 & PKO2 & PKO3 & GL-97 & NL1 & NL3 & NLSH & TM1 & PK1 & PKDD\\
\hline
 $M^{\rm DU}~[M_\odot]$ & 1.20 & 1.45 & 1.43 & 1.10 & 0.75 & 1.01 & 1.20 & 0.96 & 0.94 & 1.26\\
 $\rho^{\rm DU}(0)~[{\rm fm^{-3}}]$ & 0.28 & 0.33 & 0.31 & 0.33 & 0.20 & 0.23 & 0.24 & 0.24 & 0.23 & 0.33\\
\hline \hline
 $M^{\rm DU}\geqslant1.5~M_\odot$ & $-$ & $-$ & $-$ & $-$ & $-$ & $-$ & $-$ & $-$ & $-$ & $-$\\
 $M^{\rm DU}\geqslant1.35~M_\odot$ & $-$ & $+$ & $+$ & $-$ & $-$ & $-$ & $-$ & $-$ & $-$ & $-$\\
\hline\hline
 \end{tabular}
 \label{Tab:cooling}
\end{table}

\subsubsection{Pressure and maximum mass of neutron star}

\begin{figure}[htbp]
 \includegraphics[width=0.48\textwidth]{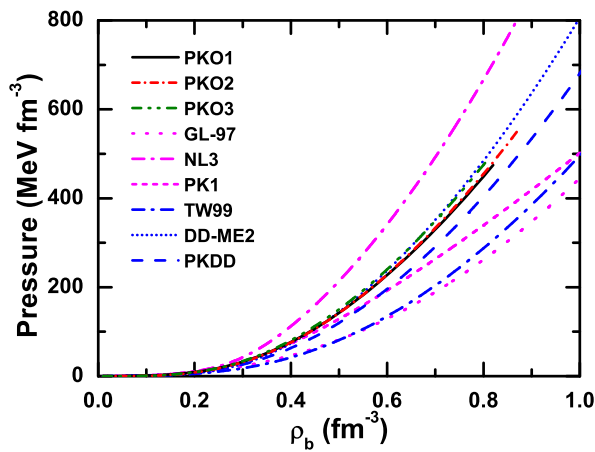}
\caption{(Color online) The pressure of neutron star matter as a
function of the baryon density $\rho_b$ (fm$^{-3}$). The results are
calculated by DDRHF with PKO1, PKO2, and PKO3, in comparison to
those by RMF with GL-97, NL3, PK1, TW99, DD-ME2, and PKDD.}
 \label{Fig:Pressure}
\end{figure}

In Fig. \ref{Fig:Pressure}, the pressures of neutron star matter
calculated by DDRHF effective interactions are shown as functions of
the baryonic density $\rho_b$. The results with RMF ones GL-97, NL3,
TW99 in Ref. \cite{Ban:2004} and also PK1, DD-ME2, PKDD have been
included for comparison. It is found that PKO1, PKO2 and PKO3
provide idential behaviors with each another over the density
dependence of the pressure, which are also close to the behaviors
predicted by RMF with PKDD and DD-ME2. Among all the DDRHF and RMF
calculations, NL3 provides the strongest density dependence and the
softest are presented by GL-97. The behaviors given by PK1 and TW99
lie between the results of DDRHF with PKO series and RMF with GL-97.

\begin{figure}[htbp]
 \includegraphics[width=0.48\textwidth]{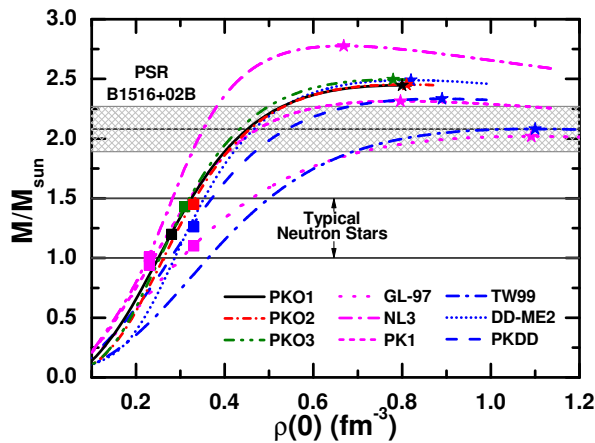}
\caption{(Color online) Neutron star mass as a function of the
central density for different DDRHF and RMF effective interactions.
Filled stars denote the maximum mass configurations, filled squares
mark the critical mass $M^{\rm DU}$ and central density values
$\rho^{\rm DU}(0)$ where the DU cooling process becomes possible.
The light grey horizontal bands around $2.08~M_\odot$ denote the
$1\sigma$ confidence level for the mass measurement of PSR
B1516$+$02B \cite{Freire:2007}. The mass region of typical neutron
stars is between $1.0~M_\odot$ and $1.5~M_\odot$.}
 \label{Fig:Mass}
\end{figure}

The variation of the pressure with respect to density is essential
to understand the structure of neutron stars. Stronger density
dependence of the pressure at high densities would lead to larger
value of maximum mass for neutron stars that can be sustained
against collapse. In Fig. \ref{Fig:Mass}, the neutron star masses
calculated by DDRHF with PKO1, PKO2, and PKO3 are shown as functions
of the central density $\rho(0)$. For comparison, are also shown the
results calculated by RMF with GL-97, NL3, PK1, TW99, DD-ME2, and
PKDD, and one could refer to Ref. \cite{Ban:2004} for more studies
with a variety of RMF effective interactions. From Fig.
\ref{Fig:Mass}, it is found that the maximum masses given by the
DDRHF calculations lie between 2.4 $M_\odot$ and 2.5 $M_\odot$ with
the central densities around 0.80 fm$^{-3}$, which are close to the
prediction of RMF with DD-ME2. Notice that these values are also
compatible to the observational constraint ($M=2.08 \pm
0.19~M_\odot$) from PSR B1516+02B \cite{Freire:2007}. In Table
\ref{Tab:mr} are shown the maximum mass limits $M_{\rm max}$ and the
corresponding central densities $\rho_{\rm max}(0)$ extracted from
Fig. \ref{Fig:Mass}. As consistent with the description of the
pressure, the non-linear RMF effective interaction NL3 presents a
rather large value of the maximum mass $M_{\rm max}=2.78~M_\odot$
with small central density $\rho_{\rm max}(0) =0.67$ fm$^{-3}$,
while the smallest $M_{\rm max}$ and the largest $\rho_{\rm max}(0)$
are obtained by RMF with GL-97 and TW99, which gives the softest
behaviors of the pressure (see Fig. \ref{Fig:Pressure}). Seen from
Table~\ref{Tab:mr}, the values of $M_{\rm max}$ given by all the
effective interactions are in the appropriate agreements with the
constraint on the maximum mass from PSR B1516+02B.

\begin{table}[htbp] \setlength{\tabcolsep}{0.25em}
\caption{Maximum mass limits $M_{\rm max}$ ($M_\odot$), the corresponding central densities
$\rho_{\rm max}(0)$ (fm$^{-3}$) and radii $R(M_{\rm max})$ (km) for neutron stars calculated by
DDRHF and RMF effective interactions. The radii (km) for 1.4 $M_\odot$ neutron stars are shown as
well.} \label{Tab:mr}
\begin{tabular}{c|*{3}{c}|*{6}{c}|*{4}{c}}\hline\hline
                    & PKO1& PKO2& PKO3&GL-97&  NL1& NL3 & NLSH& TM1 & PK1 & TW99&DD-ME1&DD-ME2& PKDD\\ \hline
 $M_{\rm max}$      & 2.45& 2.45& 2.49& 2.02& 2.81& 2.78& 2.80& 2.18& 2.32& 2.08& 2.45 & 2.49 & 2.33\\
 $\rho_{\rm max}(0)$& 0.80& 0.81& 0.78& 1.09& 0.66& 0.67& 0.65& 0.85& 0.80& 1.10& 0.84 & 0.82 & 0.89\\
 $R(M_{\rm max})$   & 12.4& 12.3& 12.5& 10.9& 13.4& 13.3& 13.5& 12.4& 12.7& 10.7& 11.9 & 12.1 & 11.8\\ \hline
 $R(1.4M_\odot)$    & 14.1& 13.8& 13.9& 13.3& 14.7& 14.7& 14.9& 14.4& 14.5& 12.4& 13.2 & 13.3 & 13.7\\ \hline\hline
\end{tabular}
\end{table}

\subsubsection{Mass-Radius relation and observational constraint}

Recent astronomic observations also provide the constraints on the
mass-radius relation of neutron stars. In this paper, four typical
observations are adopted to test the theoretical calculations.
 \begin{enumerate}
\item The large radiation radius $R_\infty=16.8$ km ($R_\infty=R/\sqrt{1-2GM/Rc^2}$) from the
isolated neutron star RX J1856 \cite{Trumper:2003we}.
\item The redshift $z\simeq0.345$, the mass $M\ge 2.10\pm 0.28~M_\odot$ and the radius
$R \ge 13.8 \pm 1.8$ km constraints in LMXBs EXO 0748-676
\cite{Cottam:2002, Ozel:2006km}.
\item $M\lesssim1.8~M_\odot$ and $R\lesssim15~{\rm km}$ constraints from the
highest frequency of QPOs 1330 Hz ever observed in 4U 0614+09
\cite{Miller04}.
\item Several neutron stars in LMXBs have gravitational masses between $1.9~M_\odot$ and possibly
$2.1~M_\odot$ from the QPOs data analysis in LMXBs 4U 1636-536 \cite{Barret:2005wd}.
\end{enumerate}

\begin{figure}[htbp]
 \includegraphics[width=0.48\textwidth]{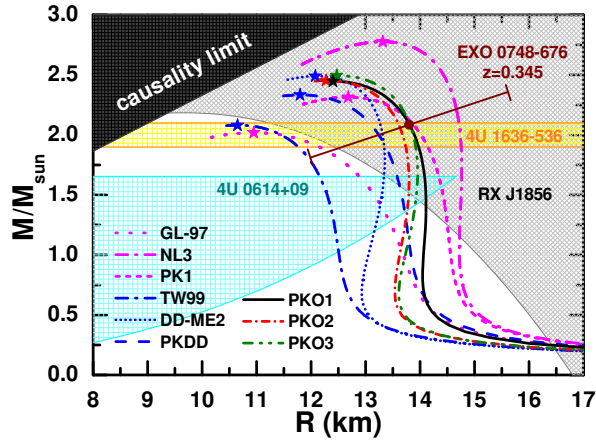}
\caption{(Color online) Mass-radius relations of neutron stars
provided by the DDRHF and RMF calculations and the corresponding
maximum masses are marked in filled star symbols. For comparison,
are also shown the four separate observational constraints from RX
J1856 (gray grided region), 4U 0614+09 (cyan grided area), 4U
1636-536 (yellow grided area) and EXO 0748-676 (wine line for
$1\sigma$ error). The black region is excluded by causality that
$R>2.9GM/c^2$ \cite{Causality1, Glendenning:1992}. See the text for
details.} \label{Fig:Radius}
\end{figure}

In Fig. \ref{Fig:Radius} are shown the mass-radius relations of
neutron stars calculated by DDRHF with PKO1, PKO2 and PKO3, and RMF
with GL-97, NL3, PK1, TW99, DD-ME2 and PKDD. The results with more
RMF effective interactions have been investigated in Ref.
\cite{Ban:2004}. For comparison, the selected observational
constraints are marked with different colors and grids as shwon in
Fig. \ref{Fig:Radius}. The causality limit that $\sqrt{\partial
p/\partial \varepsilon}\leq 1$ results in $R>2.9GM/c^2$
\cite{Causality1, Glendenning:1992} and the corresponding region in
Fig. \ref{Fig:Radius} is maked in black. Compared to all the
observational limits, it is found that better agreements are
obtained by the DDRHF effective interactions than the RMF ones.
Among the RMF results, GL-97 is excluded by the limits from RX
J1856, and TW99 is excluded by the limits from both RX J1856 and EXO
0748-676, while NL3 could not fulfil the constraint from 4U 0614+09.
If upper mass limit $2.1~M_\odot$ is taken in 4U 1636-536, GL-97 and
TW99 (just a marginal cover) is not satisfied either. The detailed
criterions of the M-R constraints are presented in Table
\ref{tab:criterion}. It is shown that the predictions given by DDRHF
with PKO series and RMF with PK1, TM1, DD-ME1, DD-ME2 and PKDD
fulfill all the M-R constraints.

\begin{table}[htbp] \setlength{\tabcolsep}{0.25em}
\caption{The criterion of the M-R constraints: (1) the isolated
neutron star RX J1856, (2) EXO 0748-676, (3) the low-mass X-ray
binary 4U 0614+09, (4-u) 4U 1636-536 with its upper mass limits, and
(4-l) 4U 1636-536 with its lower mass limits. Fulfillment
(violation) of a constraint is indicated with $+(-)$ and the
marginal cover is marked with $\vartriangle$. See the text for
details.} \label{tab:criterion}
 \begin{tabular}{c|*{3}{c}|*{6}{c}|*{4}{c}}
\hline\hline
     & PKO1 & PKO2 & PKO3 & GL-97 & NL1 & NL3 & NLSH & TM1 & PK1 & TW99 & DD-ME1 & DD-ME2 & PKDD\\ \hline
 1   & $+$ & $+$ & $+$ & $-$ & $+$ & $+$ & $+$ & $+$ & $+$ & $-$ & $+$ & $+$ & $+$\\
 2   & $+$ & $+$ & $+$ & $+$ & $+$ & $+$ & $+$ & $+$ & $+$ & $\vartriangle$ & $+$ & $+$ & $+$\\
 3   & $+$ & $+$ & $+$ & $+$ & $\vartriangle$ & $\vartriangle$ & $-$ & $+$ & $+$ & $+$ & $+$ & $+$ & $+$\\
 4-u & $+$ & $+$ & $+$ & $-$ & $+$ & $+$ & $+$ & $+$ & $+$ & $\vartriangle$ & $+$ & $+$ & $+$\\
 4-l & $+$ & $+$ & $+$ & $+$ & $+$ & $+$ & $+$ & $+$ & $+$ & $+$ & $+$ & $+$ & $+$\\
\hline\hline
 \end{tabular}
\end{table}

In Refs. \cite{Horowitz:2001, Horowitz:2001PRC}, the radius of
neutron stars with the mass 1.4 $M_\odot$ was found to be correlated
with the neutron skin thickness of $^{208}$Pb as well as the
symmetry energy. If the oberservation can limit the radius of
neutron stars to a narrow range, a strong constraint can be imposed
on the symmetry energy. On the other hand, if the neutron skin
thickness of $^{208}$Pb or the symmetry energy could be precisely
determined from the terrestrial experiments, it will be helpful to
understand the neutron star structure and rule out some EoSs for the
neutron star matter. As seen in Fig. \ref{Fig:Radius} and Table
\ref{Tab:mr}, although several equations of state provide similar
maximum masses, they still show some discrepancy for the radius of
neutron stars with the mass 1.4 $M_\odot$. In Table \ref{Tab:mr},
the DDRHF interactions predict this radius in a range from 13.8 km
to 14.1 km, while the nonlinear RMF one NLSH gives the largest value
14.9 km, and the density dependent RMF one TW99 shows the smallest
12.4 km. All the calculated results except TW99 are coincident with
the X-ray spectral analysis of the quiescent LMXB X7 in the globular
cluster 47 Tuc, which requires a rather large radius of
$14.5^{+1.8}_{-1.6}$ km for 1.4 $M_\odot$ compact stars
\cite{Heinke:2006}.

\section{Summary}\label{sec:sum}

In this paper, the equations of state for symmetric nuclear matter,
pure neutron matter and $\beta$-stable neutron star matter have been
studied within the density dependent relativistic Hartree-Fock
(DDRHF) theory with PKO1, PKO2 and PKO3. Substantial effects from
the Fock terms are found in describing the asymmetric nuclear matter
at high densities. Due to the contributions from the Fock terms of
$\sigma$- and $\omega$-couplings, stronger density dependence on the
symmetry energy is obtained from DDRHF at high densities, as
compared to the RMF calculations with the density dependent
meson-nucleon couplings. Because of the weak density dependence of
$g_\rho$ in PKO1, which induces remarkable contributions from the
$\rho$-meson coupling to the symmetry energy, PKO1 shows stronger
density dependence on the symmetry energy than both PKO2 and PKO3.
With the obtained equations of state for $\beta$-stable nuclear
matter, the properties of neutron stars are investigated within the
DDRHF theory for the first time and the recent observational
constraints of compact stars are also introduced to test the
applicability of the DDRHF models.

Due to the extra enhancement from the $\sigma$ and $\omega$ exchange
terms on the symmetry energy, large proton fractions in neutron
stars are predicted by the DDRHF calculations, which affects
essentially the cooling process of the star. For the DU process
occurring, DDRHF with PKO2 and PKO3 gives critical neutron star mass
$\sim 1.45~M_\odot$, which are close to the limit 1.5 $M_\odot$ from
the modern soft X-ray data analysis in the temperature-age diagram
and fulfil the weaker constraint $1.35~M_\odot$. In contrast, fairly
small mass limits are presented by the calculations of DDRHF with
PKO1, RMF with the non-linear self-couplings of mesons, and RMF with
PKDD, mainly due to their stronger $\rho$-coupling contributions to
the symmetry energy at high densities. Different from these two
cases, the RMF calculations with TW99, DD-ME1 and DD-ME2 do not
support the occurrence of DU process in neutron stars at all. In
addition, the radii of 1.4 $M_\odot$ neutron stars are correlated
with the symmetry energy as well. In general, stronger density
dependence on the symmetry energy leads to larger radius for 1.4
$M_\odot$ neutron star. The radii given by the DDRHF calculations
lie between 13.8 and 14.1 km, larger than the RMF calculations with
the density-dependent meson-nucleon couplings, and smaller than the
ones with the non-linear self-couplings of mesons except GL-97.

For the maximum masses and central densities of neutron stars, they
are tightly correlated with the behavior of the pressure with
respect to the density. Due to the similar density dependent
behaviors of the pressure, identical maximum masses
($\sim2.5~M_\odot$) of neutron stars are found in the calculations
of DDRHF, and RMF with DD-ME1 and DD-ME2, as well as the central
densities around 0.80 fm$^{-3}$. The results are in reasonable
agreement with high pulsar mass $2.08 \pm 0.19~M_\odot$ from PSR
B1516+02B recently reported. The mass-radius relations of neutron
stars determined by the DDRHF calculations are also consistent with
the observational data from thermal radiation measurement in the
isolated neutron star RX J1856, QPOs frequency limits in LMXBs 4U
0614+09 and 4U 1636-536, and the redshift limit determined in LMXBs
EXO 0748-676, which are only partially satisfied in the RMF
calculations with GL-97, NL1, NL3, NLSH and TW99.

\begin{acknowledgements}
The authors thank H.-J. Schulze for the stimulating discussions and
D. Blaschke for communications about the observational constraints
of compact stars. This work is partly supported by Major State Basic
Research Development Program (2007CB815000), the National Natural
Science Foundation of China (10435010, 10775004, and 10221003) and
Asia-Europe Link Program in Nuclear Physics and Astrophysics
(CN/ASIA-LINK/008 094-791).
\end{acknowledgements}


\begin{thebibliography}{79}
\expandafter\ifx\csname
natexlab\endcsname\relax\def\natexlab#1{#1}\fi
\expandafter\ifx\csname bibnamefont\endcsname\relax
  \def\bibnamefont#1{#1}\fi
\expandafter\ifx\csname bibfnamefont\endcsname\relax
  \def\bibfnamefont#1{#1}\fi
\expandafter\ifx\csname citenamefont\endcsname\relax
  \def\citenamefont#1{#1}\fi
\expandafter\ifx\csname url\endcsname\relax
  \def\url#1{\texttt{#1}}\fi
\expandafter\ifx\csname urlprefix\endcsname\relax\def\urlprefix{URL
}\fi \providecommand{\bibinfo}[2]{#2}
\providecommand{\eprint}[2][]{\url{#2}}

\bibitem[{\citenamefont{Weber}(1999)}]{BookWeber99}
\bibinfo{author}{\bibfnamefont{F.}~\bibnamefont{Weber}},
  \emph{\bibinfo{title}{Pulsars as Astrophysical Laboratories for Nuclear and
  Particle Physics}} (\bibinfo{publisher}{IOP}, \bibinfo{address}{Bristol},
  \bibinfo{year}{1999}).

\bibitem[{\citenamefont{Shapiro and Teukolsky}(1983)}]{Shapiro83}
\bibinfo{author}{\bibfnamefont{S.~L.} \bibnamefont{Shapiro}} \bibnamefont{and}
  \bibinfo{author}{\bibfnamefont{S.~A.} \bibnamefont{Teukolsky}},
  \emph{\bibinfo{title}{Black Holes, White Dwarfs, and Neutron Stars, The
  Physics of the Compact Objects}} (\bibinfo{publisher}{Wiley},
  \bibinfo{address}{New York}, \bibinfo{year}{1983}).

\bibitem[{\citenamefont{Weber et~al.}(2007)\citenamefont{Weber, Negreiros,
  Rosenfield, and Stejner}}]{Weber:2007}
\bibinfo{author}{\bibfnamefont{F.}~\bibnamefont{Weber}},
  \bibinfo{author}{\bibfnamefont{R.}~\bibnamefont{Negreiros}},
  \bibinfo{author}{\bibfnamefont{P.}~\bibnamefont{Rosenfield}},
  \bibnamefont{and} \bibinfo{author}{\bibfnamefont{M.}~\bibnamefont{Stejner}},
  \bibinfo{journal}{Prog. Part. Nucl. Phys.} \textbf{\bibinfo{volume}{59}},
  \bibinfo{pages}{94} (\bibinfo{year}{2007}).

\bibitem[{\citenamefont{Danielewicz et~al.}(2002)\citenamefont{Danielewicz,
  Lacey, and Lynch}}]{Danielewicz:2002}
\bibinfo{author}{\bibfnamefont{P.}~\bibnamefont{Danielewicz}},
  \bibinfo{author}{\bibfnamefont{R.}~\bibnamefont{Lacey}}, \bibnamefont{and}
  \bibinfo{author}{\bibfnamefont{W.~G.} \bibnamefont{Lynch}},
  \bibinfo{journal}{Science} \textbf{\bibinfo{volume}{298}},
  \bibinfo{pages}{1592} (\bibinfo{year}{2002}).

\bibitem[{\citenamefont{Fuchs}(2006)}]{Fuchs:2006Reports}
\bibinfo{author}{\bibfnamefont{C.}~\bibnamefont{Fuchs}},
  \bibinfo{journal}{Phys. Rep.} \textbf{\bibinfo{volume}{56}},
  \bibinfo{pages}{1} (\bibinfo{year}{2006}).

\bibitem[{\citenamefont{Li et~al.}(2008)\citenamefont{Li, Chen, and
  Ko}}]{BALi:2008Reports}
\bibinfo{author}{\bibfnamefont{B.-A.} \bibnamefont{Li}},
  \bibinfo{author}{\bibfnamefont{L.-W.} \bibnamefont{Chen}}, \bibnamefont{and}
  \bibinfo{author}{\bibfnamefont{C.~M.} \bibnamefont{Ko}},
  \bibinfo{journal}{Phys. Rep.} \textbf{\bibinfo{volume}{464}},
  \bibinfo{pages}{113} (\bibinfo{year}{2008}).

\bibitem[{\citenamefont{Landau}(1932)}]{Landau:1932}
\bibinfo{author}{\bibfnamefont{L.}~\bibnamefont{Landau}},
  \bibinfo{journal}{Physikalische Zeitschrift der Sowjetunion}
  \textbf{\bibinfo{volume}{1}}, \bibinfo{pages}{285} (\bibinfo{year}{1932}).

\bibitem[{\citenamefont{Baade and Zwicky}(1934)}]{Baade:1934}
\bibinfo{author}{\bibfnamefont{W.}~\bibnamefont{Baade}} \bibnamefont{and}
  \bibinfo{author}{\bibfnamefont{F.}~\bibnamefont{Zwicky}},
  \bibinfo{journal}{Proc. Nat. Acad. Sci. U.S.A.}
  \textbf{\bibinfo{volume}{20}}, \bibinfo{pages}{255} (\bibinfo{year}{1934}).

\bibitem[{\citenamefont{Hewish et~al.}(1968)\citenamefont{Hewish, Bell,
  Pilkington, Scott, and Collins}}]{Hewish:1968}
\bibinfo{author}{\bibfnamefont{A.}~\bibnamefont{Hewish}},
  \bibinfo{author}{\bibfnamefont{S.~J.} \bibnamefont{Bell}},
  \bibinfo{author}{\bibfnamefont{J.~D.~H.} \bibnamefont{Pilkington}},
  \bibinfo{author}{\bibfnamefont{P.~F.} \bibnamefont{Scott}}, \bibnamefont{and}
  \bibinfo{author}{\bibfnamefont{R.~A.} \bibnamefont{Collins}},
  \bibinfo{journal}{Nature} \textbf{\bibinfo{volume}{217}},
  \bibinfo{pages}{709} (\bibinfo{year}{1968}).

\bibitem[{\citenamefont{Pacini}(1967)}]{Pacini:1968}
\bibinfo{author}{\bibfnamefont{F.}~\bibnamefont{Pacini}},
  \bibinfo{journal}{Nature} \textbf{\bibinfo{volume}{216}},
  \bibinfo{pages}{567} (\bibinfo{year}{1967}).

\bibitem[{\citenamefont{Lattimer and Prakash}(2004)}]{Lattimer:2004}
\bibinfo{author}{\bibfnamefont{J.~M.} \bibnamefont{Lattimer}} \bibnamefont{and}
  \bibinfo{author}{\bibfnamefont{M.}~\bibnamefont{Prakash}},
  \bibinfo{journal}{Science} \textbf{\bibinfo{volume}{304}},
  \bibinfo{pages}{536} (\bibinfo{year}{2004}).

\bibitem[{\citenamefont{Lattimer and Prakash}(2007)}]{Lattimer:2007}
\bibinfo{author}{\bibfnamefont{J.~M.} \bibnamefont{Lattimer}} \bibnamefont{and}
  \bibinfo{author}{\bibfnamefont{M.}~\bibnamefont{Prakash}},
  \bibinfo{journal}{Phys. Rep.} \textbf{\bibinfo{volume}{442}},
  \bibinfo{pages}{109} (\bibinfo{year}{2007}).

\bibitem[{\citenamefont{Barret et~al.}(2005)\citenamefont{Barret, Olive, and
  Miller}}]{Barret:2005wd}
\bibinfo{author}{\bibfnamefont{D.}~\bibnamefont{Barret}},
  \bibinfo{author}{\bibfnamefont{J.-F.} \bibnamefont{Olive}}, \bibnamefont{and}
  \bibinfo{author}{\bibfnamefont{M.~C.} \bibnamefont{Miller}},
  \bibinfo{journal}{Mon. Not. Roy. Astron. Soc.}
  \textbf{\bibinfo{volume}{361}}, \bibinfo{pages}{855} (\bibinfo{year}{2005}).

\bibitem[{\citenamefont{Freire et~al.}(2008{\natexlab{a}})\citenamefont{Freire,
  Wolszczan, van~den Berg, and Hessels}}]{Freire:2007}
\bibinfo{author}{\bibfnamefont{P.~C.~C.} \bibnamefont{Freire}},
  \bibinfo{author}{\bibfnamefont{A.}~\bibnamefont{Wolszczan}},
  \bibinfo{author}{\bibfnamefont{M.}~\bibnamefont{van~den Berg}},
  \bibnamefont{and} \bibinfo{author}{\bibfnamefont{J.~W.~T.}
  \bibnamefont{Hessels}}, \bibinfo{journal}{Astrophys. J.}
  \textbf{\bibinfo{volume}{679}}, \bibinfo{pages}{1433}
  (\bibinfo{year}{2008}{\natexlab{a}}).

\bibitem[{\citenamefont{Freire et~al.}(2008{\natexlab{b}})\citenamefont{Freire,
  Ransom, B\'egin, Stairs, Hessels, Frey, and Camilo}}]{Freire:2008}
\bibinfo{author}{\bibfnamefont{P.~C.~C.} \bibnamefont{Freire}},
  \bibinfo{author}{\bibfnamefont{S.~M.} \bibnamefont{Ransom}},
  \bibinfo{author}{\bibfnamefont{S.}~\bibnamefont{B\'egin}},
  \bibinfo{author}{\bibfnamefont{I.~H.} \bibnamefont{Stairs}},
  \bibinfo{author}{\bibfnamefont{J.~W.~T.} \bibnamefont{Hessels}},
  \bibinfo{author}{\bibfnamefont{L.~H.} \bibnamefont{Frey}}, \bibnamefont{and}
  \bibinfo{author}{\bibfnamefont{F.}~\bibnamefont{Camilo}},
  \bibinfo{journal}{Astrophys. J.} \textbf{\bibinfo{volume}{675}},
  \bibinfo{pages}{670} (\bibinfo{year}{2008}{\natexlab{b}}).

\bibitem[{\citenamefont{Tr\"umper et~al.}(2004)\citenamefont{Tr\"umper,
  Burwitz, Haberl, and Zavlin}}]{Trumper:2003we}
\bibinfo{author}{\bibfnamefont{J.~E.} \bibnamefont{Tr\"umper}},
  \bibinfo{author}{\bibfnamefont{V.}~\bibnamefont{Burwitz}},
  \bibinfo{author}{\bibfnamefont{F.}~\bibnamefont{Haberl}}, \bibnamefont{and}
  \bibinfo{author}{\bibfnamefont{V.~E.} \bibnamefont{Zavlin}},
  \bibinfo{journal}{Nucl. Phys. Proc. Suppl.} \textbf{\bibinfo{volume}{132}},
  \bibinfo{pages}{560} (\bibinfo{year}{2004}).

\bibitem[{\citenamefont{Heinke et~al.}(2006)\citenamefont{Heinke, Rybicki,
  Narayan, and Grindlay}}]{Heinke:2006}
\bibinfo{author}{\bibfnamefont{C.~O.} \bibnamefont{Heinke}},
  \bibinfo{author}{\bibfnamefont{G.~B.} \bibnamefont{Rybicki}},
  \bibinfo{author}{\bibfnamefont{R.}~\bibnamefont{Narayan}}, \bibnamefont{and}
  \bibinfo{author}{\bibfnamefont{J.~E.} \bibnamefont{Grindlay}},
  \bibinfo{journal}{Astrophys. J.} \textbf{\bibinfo{volume}{644}},
  \bibinfo{pages}{1090} (\bibinfo{year}{2006}).

\bibitem[{\citenamefont{Cottam et~al.}(2002)\citenamefont{Cottam, Paerls, and
  Mendez}}]{Cottam:2002}
\bibinfo{author}{\bibfnamefont{J.}~\bibnamefont{Cottam}},
  \bibinfo{author}{\bibfnamefont{F.}~\bibnamefont{Paerls}}, \bibnamefont{and}
  \bibinfo{author}{\bibfnamefont{M.}~\bibnamefont{Mendez}},
  \bibinfo{journal}{Nature} \textbf{\bibinfo{volume}{420}}, \bibinfo{pages}{51}
  (\bibinfo{year}{2002}).

\bibitem[{\citenamefont{\"Ozel}(2006)}]{Ozel:2006km}
\bibinfo{author}{\bibfnamefont{F.}~\bibnamefont{\"Ozel}},
  \bibinfo{journal}{Nature} \textbf{\bibinfo{volume}{441}},
  \bibinfo{pages}{1115} (\bibinfo{year}{2006}).

\bibitem[{\citenamefont{Miller}(2004)}]{Miller04}
\bibinfo{author}{\bibfnamefont{M.~C.} \bibnamefont{Miller}},
  \bibinfo{journal}{AIP Conf. Proc.} \textbf{\bibinfo{volume}{714}},
  \bibinfo{pages}{365} (\bibinfo{year}{2004}).

\bibitem[{\citenamefont{Lattimer et~al.}(1991)\citenamefont{Lattimer, Pethick,
  Prakash, and Haensel}}]{Lattimer:1991}
\bibinfo{author}{\bibfnamefont{J.~M.} \bibnamefont{Lattimer}},
  \bibinfo{author}{\bibfnamefont{C.~J.} \bibnamefont{Pethick}},
  \bibinfo{author}{\bibfnamefont{M.}~\bibnamefont{Prakash}}, \bibnamefont{and}
  \bibinfo{author}{\bibfnamefont{P.}~\bibnamefont{Haensel}},
  \bibinfo{journal}{Phys. Rev. Lett.} \textbf{\bibinfo{volume}{66}},
  \bibinfo{pages}{2701} (\bibinfo{year}{1991}).

\bibitem[{\citenamefont{Blaschke et~al.}(2004)\citenamefont{Blaschke,
  Grigorian, and Voskresensky}}]{Blaschke:2004}
\bibinfo{author}{\bibfnamefont{D.}~\bibnamefont{Blaschke}},
  \bibinfo{author}{\bibfnamefont{H.}~\bibnamefont{Grigorian}},
  \bibnamefont{and} \bibinfo{author}{\bibfnamefont{D.~N.}
  \bibnamefont{Voskresensky}}, \bibinfo{journal}{Astron. Astrophys.}
  \textbf{\bibinfo{volume}{424}}, \bibinfo{pages}{979} (\bibinfo{year}{2004}).

\bibitem[{\citenamefont{Popov et~al.}(2006)\citenamefont{Popov, Grigorian,
  Turolla, and Blaschke}}]{Popov:2006}
\bibinfo{author}{\bibfnamefont{S.}~\bibnamefont{Popov}},
  \bibinfo{author}{\bibfnamefont{H.}~\bibnamefont{Grigorian}},
  \bibinfo{author}{\bibfnamefont{R.}~\bibnamefont{Turolla}}, \bibnamefont{and}
  \bibinfo{author}{\bibfnamefont{D.}~\bibnamefont{Blaschke}},
  \bibinfo{journal}{Astron. Astrophys.} \textbf{\bibinfo{volume}{448}},
  \bibinfo{pages}{327} (\bibinfo{year}{2006}).

\bibitem[{\citenamefont{Kl\"ahn et~al.}(2006)\citenamefont{Kl\"ahn, Blaschke,
  Typel, van Dalen, Faessler, Fuchs, Gaitanos, Grigorian, Ho, Kolomeitsev
  et~al.}}]{Klahn:2006ir}
\bibinfo{author}{\bibfnamefont{T.}~\bibnamefont{Kl\"ahn}},
  \bibinfo{author}{\bibfnamefont{D.}~\bibnamefont{Blaschke}},
  \bibinfo{author}{\bibfnamefont{S.}~\bibnamefont{Typel}},
  \bibinfo{author}{\bibfnamefont{E.~N.~E.} \bibnamefont{van Dalen}},
  \bibinfo{author}{\bibfnamefont{A.}~\bibnamefont{Faessler}},
  \bibinfo{author}{\bibfnamefont{C.}~\bibnamefont{Fuchs}},
  \bibinfo{author}{\bibfnamefont{T.}~\bibnamefont{Gaitanos}},
  \bibinfo{author}{\bibfnamefont{H.}~\bibnamefont{Grigorian}},
  \bibinfo{author}{\bibfnamefont{A.}~\bibnamefont{Ho}},
  \bibinfo{author}{\bibfnamefont{E.~E.} \bibnamefont{Kolomeitsev}},
  \bibnamefont{et~al.}, \bibinfo{journal}{Phys. Rev.}
  \textbf{\bibinfo{volume}{C 74}}, \bibinfo{pages}{035802}
  (\bibinfo{year}{2006}).

\bibitem[{\citenamefont{Miller and Green}(1972)}]{Miller:1972}
\bibinfo{author}{\bibfnamefont{L.~D.} \bibnamefont{Miller}} \bibnamefont{and}
  \bibinfo{author}{\bibfnamefont{A.~E.~S.} \bibnamefont{Green}},
  \bibinfo{journal}{Phys. Rev.} \textbf{\bibinfo{volume}{C 5}},
  \bibinfo{pages}{241} (\bibinfo{year}{1972}).

\bibitem[{\citenamefont{Walecka}(1974)}]{Walecka:1974}
\bibinfo{author}{\bibfnamefont{J.~D.} \bibnamefont{Walecka}},
  \bibinfo{journal}{Ann. Phys. (N.Y.)} \textbf{\bibinfo{volume}{83}},
  \bibinfo{pages}{491} (\bibinfo{year}{1974}).

\bibitem[{\citenamefont{Serot and Walecka}(1986)}]{Serot:1986}
\bibinfo{author}{\bibfnamefont{B.}~\bibnamefont{Serot}} \bibnamefont{and}
  \bibinfo{author}{\bibfnamefont{J.~D.} \bibnamefont{Walecka}},
  \bibinfo{journal}{Adv. Nucl. Phys.} \textbf{\bibinfo{volume}{16}},
  \bibinfo{pages}{1} (\bibinfo{year}{1986}).

\bibitem[{\citenamefont{Reinhard}(1989)}]{Reinhard:1989}
\bibinfo{author}{\bibfnamefont{P.~G.} \bibnamefont{Reinhard}},
  \bibinfo{journal}{Reports on Progress in Physics}
  \textbf{\bibinfo{volume}{52}}, \bibinfo{pages}{439} (\bibinfo{year}{1989}).

\bibitem[{\citenamefont{Ring}(1996)}]{Ring:1996}
\bibinfo{author}{\bibfnamefont{P.}~\bibnamefont{Ring}}, \bibinfo{journal}{Prog.
  Part. Nucl. Phys.} \textbf{\bibinfo{volume}{37}}, \bibinfo{pages}{193}
  (\bibinfo{year}{1996}).

\bibitem[{\citenamefont{Serot and Walecka}(1997)}]{Serot:1997}
\bibinfo{author}{\bibfnamefont{B.~D.} \bibnamefont{Serot}} \bibnamefont{and}
  \bibinfo{author}{\bibfnamefont{J.~D.} \bibnamefont{Walecka}},
  \bibinfo{journal}{Int. J. Mod. Phys.} \textbf{\bibinfo{volume}{E 6}},
  \bibinfo{pages}{515} (\bibinfo{year}{1997}).

\bibitem[{\citenamefont{Bender et~al.}(2003)\citenamefont{Bender, Heenen, and
  Reinhard}}]{Bender:2003}
\bibinfo{author}{\bibfnamefont{M.}~\bibnamefont{Bender}},
  \bibinfo{author}{\bibfnamefont{P.-H.} \bibnamefont{Heenen}},
  \bibnamefont{and} \bibinfo{author}{\bibfnamefont{P.-G.}
  \bibnamefont{Reinhard}}, \bibinfo{journal}{Revs. Mod. Phys.}
  \textbf{\bibinfo{volume}{75}}, \bibinfo{pages}{121} (\bibinfo{year}{2003}).

\bibitem[{\citenamefont{Meng}(1998)}]{meng98npa}
\bibinfo{author}{\bibfnamefont{J.}~\bibnamefont{Meng}}, \bibinfo{journal}{Nucl.
  Phys.} \textbf{\bibinfo{volume}{A 635}}, \bibinfo{pages}{3}
  (\bibinfo{year}{1998}).

\bibitem[{\citenamefont{Meng and Ring}(1996)}]{meng96prl}
\bibinfo{author}{\bibfnamefont{J.}~\bibnamefont{Meng}} \bibnamefont{and}
  \bibinfo{author}{\bibfnamefont{P.}~\bibnamefont{Ring}},
  \bibinfo{journal}{Phys. Rev. Lett.} \textbf{\bibinfo{volume}{77}},
  \bibinfo{pages}{3963} (\bibinfo{year}{1996}).

\bibitem[{\citenamefont{Meng and Ring}(1998)}]{meng98prl}
\bibinfo{author}{\bibfnamefont{J.}~\bibnamefont{Meng}} \bibnamefont{and}
  \bibinfo{author}{\bibfnamefont{P.}~\bibnamefont{Ring}},
  \bibinfo{journal}{Phys. Rev. Lett.} \textbf{\bibinfo{volume}{80}},
  \bibinfo{pages}{460} (\bibinfo{year}{1998}).

\bibitem[{\citenamefont{Meng et~al.}(1998)\citenamefont{Meng, Tanihata, and
  Yamaji}}]{meng98plb}
\bibinfo{author}{\bibfnamefont{J.}~\bibnamefont{Meng}},
  \bibinfo{author}{\bibfnamefont{I.}~\bibnamefont{Tanihata}}, \bibnamefont{and}
  \bibinfo{author}{\bibfnamefont{S.}~\bibnamefont{Yamaji}},
  \bibinfo{journal}{Phys. Lett.} \textbf{\bibinfo{volume}{B 419}},
  \bibinfo{pages}{1} (\bibinfo{year}{1998}).

\bibitem[{\citenamefont{Meng et~al.}(2006)\citenamefont{Meng, Toki, Zhou,
  Zhang, Long, and Geng}}]{Meng:2006}
\bibinfo{author}{\bibfnamefont{J.}~\bibnamefont{Meng}},
  \bibinfo{author}{\bibfnamefont{H.}~\bibnamefont{Toki}},
  \bibinfo{author}{\bibfnamefont{S.~G.} \bibnamefont{Zhou}},
  \bibinfo{author}{\bibfnamefont{S.~Q.} \bibnamefont{Zhang}},
  \bibinfo{author}{\bibfnamefont{W.~H.} \bibnamefont{Long}}, \bibnamefont{and}
  \bibinfo{author}{\bibfnamefont{L.~S.} \bibnamefont{Geng}},
  \bibinfo{journal}{Prog. Part. Nucl. Phys.} \textbf{\bibinfo{volume}{57}},
  \bibinfo{pages}{470} (\bibinfo{year}{2006}).

\bibitem[{\citenamefont{Lalazissis et~al.}(1997)\citenamefont{Lalazissis,
  K\"onig, and Ring}}]{Lalazissis:1997}
\bibinfo{author}{\bibfnamefont{G.~A.} \bibnamefont{Lalazissis}},
  \bibinfo{author}{\bibfnamefont{J.}~\bibnamefont{K\"onig}}, \bibnamefont{and}
  \bibinfo{author}{\bibfnamefont{P.}~\bibnamefont{Ring}},
  \bibinfo{journal}{Phys. Rev.} \textbf{\bibinfo{volume}{C 55}},
  \bibinfo{pages}{540} (\bibinfo{year}{1997}).

\bibitem[{\citenamefont{Typel and Wolter}(1999)}]{Typel:1999}
\bibinfo{author}{\bibfnamefont{S.}~\bibnamefont{Typel}} \bibnamefont{and}
  \bibinfo{author}{\bibfnamefont{H.~H.} \bibnamefont{Wolter}},
  \bibinfo{journal}{Nucl. Phys.} \textbf{\bibinfo{volume}{A 656}},
  \bibinfo{pages}{331} (\bibinfo{year}{1999}).

\bibitem[{\citenamefont{Long et~al.}(2004)\citenamefont{Long, Meng, VanGiai,
  and Zhou}}]{Long04}
\bibinfo{author}{\bibfnamefont{W.}~\bibnamefont{Long}},
  \bibinfo{author}{\bibfnamefont{J.}~\bibnamefont{Meng}},
  \bibinfo{author}{\bibfnamefont{N.}~\bibnamefont{VanGiai}}, \bibnamefont{and}
  \bibinfo{author}{\bibfnamefont{S.-G.} \bibnamefont{Zhou}},
  \bibinfo{journal}{Phys. Rev.} \textbf{\bibinfo{volume}{C 69}},
  \bibinfo{pages}{034319} (\bibinfo{year}{2004}).

\bibitem[{\citenamefont{Nik\v{s}i\'{c}
  et~al.}(2002)\citenamefont{Nik\v{s}i\'{c}, Vretenar, Finelli, and
  Ring}}]{Niksic:2002}
\bibinfo{author}{\bibfnamefont{T.}~\bibnamefont{Nik\v{s}i\'{c}}},
  \bibinfo{author}{\bibfnamefont{D.}~\bibnamefont{Vretenar}},
  \bibinfo{author}{\bibfnamefont{P.}~\bibnamefont{Finelli}}, \bibnamefont{and}
  \bibinfo{author}{\bibfnamefont{P.}~\bibnamefont{Ring}},
  \bibinfo{journal}{Phys. Rev.} \textbf{\bibinfo{volume}{C 66}},
  \bibinfo{pages}{024306} (\bibinfo{year}{2002}).

\bibitem[{\citenamefont{Lalazissis et~al.}(2005)\citenamefont{Lalazissis,
  Nik\v{s}i\'{c}, Vretenar, and Ring}}]{Lalazissis:2005}
\bibinfo{author}{\bibfnamefont{G.~A.} \bibnamefont{Lalazissis}},
  \bibinfo{author}{\bibfnamefont{T.}~\bibnamefont{Nik\v{s}i\'{c}}},
  \bibinfo{author}{\bibfnamefont{D.}~\bibnamefont{Vretenar}}, \bibnamefont{and}
  \bibinfo{author}{\bibfnamefont{P.}~\bibnamefont{Ring}},
  \bibinfo{journal}{Phys. Rev.} \textbf{\bibinfo{volume}{C 71}},
  \bibinfo{pages}{024312} (\bibinfo{year}{2005}).

\bibitem[{\citenamefont{Oppenheimer and Volkoff}(1939)}]{Oppenheimer:1939}
\bibinfo{author}{\bibfnamefont{J.~R.} \bibnamefont{Oppenheimer}}
  \bibnamefont{and} \bibinfo{author}{\bibfnamefont{G.~M.}
  \bibnamefont{Volkoff}}, \bibinfo{journal}{Phys. Rev.}
  \textbf{\bibinfo{volume}{55}}, \bibinfo{pages}{374} (\bibinfo{year}{1939}).

\bibitem[{\citenamefont{Tolman}(1939)}]{Tolman:1939}
\bibinfo{author}{\bibfnamefont{R.~C.} \bibnamefont{Tolman}},
  \bibinfo{journal}{Phys. Rev.} \textbf{\bibinfo{volume}{55}},
  \bibinfo{pages}{364} (\bibinfo{year}{1939}).

\bibitem[{\citenamefont{Glendenning}(2000)}]{gle}
\bibinfo{author}{\bibfnamefont{N.~K.} \bibnamefont{Glendenning}},
  \emph{\bibinfo{title}{Compact Stars, Nuclear Physics, Particle Physics, and
  General Relativity}} (\bibinfo{publisher}{Springer-Verlag},
  \bibinfo{address}{New York}, \bibinfo{year}{2000}), \bibinfo{edition}{2nd}
  ed.

\bibitem[{\citenamefont{Brockmann and Toki}(1992)}]{Brockmann:1992}
\bibinfo{author}{\bibfnamefont{R.}~\bibnamefont{Brockmann}} \bibnamefont{and}
  \bibinfo{author}{\bibfnamefont{H.}~\bibnamefont{Toki}},
  \bibinfo{journal}{Phys. Rev. Lett.} \textbf{\bibinfo{volume}{68}},
  \bibinfo{pages}{3408} (\bibinfo{year}{1992}).

\bibitem[{\citenamefont{Lenske and Fuchs}(1995)}]{Lenske:1995}
\bibinfo{author}{\bibfnamefont{H.}~\bibnamefont{Lenske}} \bibnamefont{and}
  \bibinfo{author}{\bibfnamefont{C.}~\bibnamefont{Fuchs}},
  \bibinfo{journal}{Phys. Lett.} \textbf{\bibinfo{volume}{B 345}},
  \bibinfo{pages}{355} (\bibinfo{year}{1995}).

\bibitem[{\citenamefont{Fuchs et~al.}(1995)\citenamefont{Fuchs, Lenske, and
  Wolter}}]{Fuchs:1995}
\bibinfo{author}{\bibfnamefont{C.}~\bibnamefont{Fuchs}},
  \bibinfo{author}{\bibfnamefont{H.}~\bibnamefont{Lenske}}, \bibnamefont{and}
  \bibinfo{author}{\bibfnamefont{H.~H.} \bibnamefont{Wolter}},
  \bibinfo{journal}{Phys. Rev.} \textbf{\bibinfo{volume}{C 52}},
  \bibinfo{pages}{3043} (\bibinfo{year}{1995}).

\bibitem[{\citenamefont{Boguta and Bodmer}(1977)}]{Boguta:1977}
\bibinfo{author}{\bibfnamefont{J.}~\bibnamefont{Boguta}} \bibnamefont{and}
  \bibinfo{author}{\bibfnamefont{A.}~\bibnamefont{Bodmer}},
  \bibinfo{journal}{Nucl. Phys.} \textbf{\bibinfo{volume}{A 292}},
  \bibinfo{pages}{413} (\bibinfo{year}{1977}).

\bibitem[{\citenamefont{Sugahara and Toki}(1994)}]{Sugahara:1994}
\bibinfo{author}{\bibfnamefont{Y.}~\bibnamefont{Sugahara}} \bibnamefont{and}
  \bibinfo{author}{\bibfnamefont{H.}~\bibnamefont{Toki}},
  \bibinfo{journal}{Nucl. Phys.} \textbf{\bibinfo{volume}{A 579}},
  \bibinfo{pages}{557} (\bibinfo{year}{1994}).

\bibitem[{\citenamefont{Serot}(1979)}]{Serot:1979}
\bibinfo{author}{\bibfnamefont{B.~D.} \bibnamefont{Serot}},
  \bibinfo{journal}{Phys. Lett.} \textbf{\bibinfo{volume}{B 86}},
  \bibinfo{pages}{146} (\bibinfo{year}{1979}).

\bibitem[{\citenamefont{Sumiyoshi et~al.}(1995)\citenamefont{Sumiyoshi,
  Kuwabara, and Toki}}]{Sumiyoshi:1995}
\bibinfo{author}{\bibfnamefont{K.}~\bibnamefont{Sumiyoshi}},
  \bibinfo{author}{\bibfnamefont{H.}~\bibnamefont{Kuwabara}}, \bibnamefont{and}
  \bibinfo{author}{\bibfnamefont{H.}~\bibnamefont{Toki}},
  \bibinfo{journal}{Nucl. Phys.} \textbf{\bibinfo{volume}{A 581}},
  \bibinfo{pages}{725} (\bibinfo{year}{1995}).

\bibitem[{\citenamefont{Hofmann et~al.}(2001)\citenamefont{Hofmann, Keil, and
  Lenske}}]{Hofmann:2001}
\bibinfo{author}{\bibfnamefont{F.}~\bibnamefont{Hofmann}},
  \bibinfo{author}{\bibfnamefont{C.~M.} \bibnamefont{Keil}}, \bibnamefont{and}
  \bibinfo{author}{\bibfnamefont{H.}~\bibnamefont{Lenske}},
  \bibinfo{journal}{Phys. Rev.} \textbf{\bibinfo{volume}{C 64}},
  \bibinfo{pages}{025804} (\bibinfo{year}{2001}).

\bibitem[{\citenamefont{Ban et~al.}(2004)\citenamefont{Ban, Li, Zhang, Jia,
  Sang, and Meng}}]{Ban:2004}
\bibinfo{author}{\bibfnamefont{S.~F.} \bibnamefont{Ban}},
  \bibinfo{author}{\bibfnamefont{J.}~\bibnamefont{Li}},
  \bibinfo{author}{\bibfnamefont{S.~Q.} \bibnamefont{Zhang}},
  \bibinfo{author}{\bibfnamefont{H.~Y.} \bibnamefont{Jia}},
  \bibinfo{author}{\bibfnamefont{J.~P.} \bibnamefont{Sang}}, \bibnamefont{and}
  \bibinfo{author}{\bibfnamefont{J.}~\bibnamefont{Meng}},
  \bibinfo{journal}{Phys. Rev.} \textbf{\bibinfo{volume}{C 69}},
  \bibinfo{pages}{045805} (\bibinfo{year}{2004}).

\bibitem[{\citenamefont{Glendenning}(1982)}]{Glendenning:1982}
\bibinfo{author}{\bibfnamefont{N.~K.} \bibnamefont{Glendenning}},
  \bibinfo{journal}{Phys. Lett.} \textbf{\bibinfo{volume}{B 114}},
  \bibinfo{pages}{392} (\bibinfo{year}{1982}).

\bibitem[{\citenamefont{Glendenning}(1985)}]{gle85}
\bibinfo{author}{\bibfnamefont{N.~K.} \bibnamefont{Glendenning}},
  \bibinfo{journal}{Astrophys. J.} \textbf{\bibinfo{volume}{293}},
  \bibinfo{pages}{470} (\bibinfo{year}{1985}).

\bibitem[{\citenamefont{Knorren et~al.}(1995)\citenamefont{Knorren, Prakash,
  and Ellis}}]{Knorren:1995}
\bibinfo{author}{\bibfnamefont{R.}~\bibnamefont{Knorren}},
  \bibinfo{author}{\bibfnamefont{M.}~\bibnamefont{Prakash}}, \bibnamefont{and}
  \bibinfo{author}{\bibfnamefont{P.~J.} \bibnamefont{Ellis}},
  \bibinfo{journal}{Phys. Rev.} \textbf{\bibinfo{volume}{C 52}},
  \bibinfo{pages}{3470} (\bibinfo{year}{1995}).

\bibitem[{\citenamefont{Schaffner and Mishustin}(1996)}]{Schaffner:1996}
\bibinfo{author}{\bibfnamefont{J.}~\bibnamefont{Schaffner}} \bibnamefont{and}
  \bibinfo{author}{\bibfnamefont{I.~N.} \bibnamefont{Mishustin}},
  \bibinfo{journal}{Phys. Rev.} \textbf{\bibinfo{volume}{C 53}},
  \bibinfo{pages}{1416} (\bibinfo{year}{1996}).

\bibitem[{\citenamefont{Horowitz and
  Piekarewicz}(2001{\natexlab{a}})}]{Horowitz:2001}
\bibinfo{author}{\bibfnamefont{C.~J.} \bibnamefont{Horowitz}} \bibnamefont{and}
  \bibinfo{author}{\bibfnamefont{J.}~\bibnamefont{Piekarewicz}},
  \bibinfo{journal}{Phys. Rev. Lett.} \textbf{\bibinfo{volume}{86}},
  \bibinfo{pages}{5647} (\bibinfo{year}{2001}{\natexlab{a}}).

\bibitem[{\citenamefont{Horowitz and
  Piekarewicz}(2001{\natexlab{b}})}]{Horowitz:2001PRC}
\bibinfo{author}{\bibfnamefont{C.~J.} \bibnamefont{Horowitz}} \bibnamefont{and}
  \bibinfo{author}{\bibfnamefont{J.}~\bibnamefont{Piekarewicz}},
  \bibinfo{journal}{Phys. Rev.} \textbf{\bibinfo{volume}{C 64}},
  \bibinfo{pages}{062802(R)} (\bibinfo{year}{2001}{\natexlab{b}}).

\bibitem[{\citenamefont{Brown et~al.}(2007)\citenamefont{Brown, Shen,
  Hillhouse, Meng, and Trzci\'{n}ska}}]{Brown:2007}
\bibinfo{author}{\bibfnamefont{B.~A.} \bibnamefont{Brown}},
  \bibinfo{author}{\bibfnamefont{G.}~\bibnamefont{Shen}},
  \bibinfo{author}{\bibfnamefont{G.~C.} \bibnamefont{Hillhouse}},
  \bibinfo{author}{\bibfnamefont{J.}~\bibnamefont{Meng}}, \bibnamefont{and}
  \bibinfo{author}{\bibfnamefont{A.}~\bibnamefont{Trzci\'{n}ska}},
  \bibinfo{journal}{Phys. Rev.} \textbf{\bibinfo{volume}{C 76}},
  \bibinfo{pages}{034305} (\bibinfo{year}{2007}).

\bibitem[{\citenamefont{Horowitz and Piekarewicz}(2002)}]{Horowitz:2002}
\bibinfo{author}{\bibfnamefont{C.~J.} \bibnamefont{Horowitz}} \bibnamefont{and}
  \bibinfo{author}{\bibfnamefont{J.}~\bibnamefont{Piekarewicz}},
  \bibinfo{journal}{Phys. Rev.} \textbf{\bibinfo{volume}{C 66}},
  \bibinfo{pages}{055803} (\bibinfo{year}{2002}).

\bibitem[{\citenamefont{Huber et~al.}(1994)\citenamefont{Huber, Weber, and
  Weigel}}]{Huber:1994}
\bibinfo{author}{\bibfnamefont{H.}~\bibnamefont{Huber}},
  \bibinfo{author}{\bibfnamefont{F.}~\bibnamefont{Weber}}, \bibnamefont{and}
  \bibinfo{author}{\bibfnamefont{M.~K.} \bibnamefont{Weigel}},
  \bibinfo{journal}{Phys. Rev.} \textbf{\bibinfo{volume}{50}},
  \bibinfo{pages}{R1287} (\bibinfo{year}{1994}).

\bibitem[{\citenamefont{Engvik et~al.}(1994)\citenamefont{Engvik,
  Hjorth-Jensen, Osnes, Bao, and {\O}stgaard}}]{Engvik:1994}
\bibinfo{author}{\bibfnamefont{L.}~\bibnamefont{Engvik}},
  \bibinfo{author}{\bibfnamefont{M.}~\bibnamefont{Hjorth-Jensen}},
  \bibinfo{author}{\bibfnamefont{E.}~\bibnamefont{Osnes}},
  \bibinfo{author}{\bibfnamefont{G.}~\bibnamefont{Bao}}, \bibnamefont{and}
  \bibinfo{author}{\bibfnamefont{E.}~\bibnamefont{{\O}stgaard}},
  \bibinfo{journal}{Phys. Rev. Lett.} \textbf{\bibinfo{volume}{73}},
  \bibinfo{pages}{2650} (\bibinfo{year}{1994}).

\bibitem[{\citenamefont{Krastev and Sammarruca}(2006)}]{Krastev:2006}
\bibinfo{author}{\bibfnamefont{P.~G.} \bibnamefont{Krastev}} \bibnamefont{and}
  \bibinfo{author}{\bibfnamefont{F.}~\bibnamefont{Sammarruca}},
  \bibinfo{journal}{Phys. Rev.} \textbf{\bibinfo{volume}{C 74}},
  \bibinfo{pages}{025808} (\bibinfo{year}{2006}).

\bibitem[{\citenamefont{Baldo et~al.}(1997)\citenamefont{Baldo, Bombaci, and
  Burgio}}]{BHF1}
\bibinfo{author}{\bibfnamefont{M.}~\bibnamefont{Baldo}},
  \bibinfo{author}{\bibfnamefont{I.}~\bibnamefont{Bombaci}}, \bibnamefont{and}
  \bibinfo{author}{\bibfnamefont{G.~F.} \bibnamefont{Burgio}},
  \bibinfo{journal}{Astron. Astrophys.} \textbf{\bibinfo{volume}{328}},
  \bibinfo{pages}{274} (\bibinfo{year}{1997}).

\bibitem[{\citenamefont{Zhou et~al.}(2004)\citenamefont{Zhou, Burgio, Lombardo,
  Schulze, and Zuo}}]{BHF2}
\bibinfo{author}{\bibfnamefont{X.~R.} \bibnamefont{Zhou}},
  \bibinfo{author}{\bibfnamefont{G.~F.} \bibnamefont{Burgio}},
  \bibinfo{author}{\bibfnamefont{U.}~\bibnamefont{Lombardo}},
  \bibinfo{author}{\bibfnamefont{H.-J.} \bibnamefont{Schulze}},
  \bibnamefont{and} \bibinfo{author}{\bibfnamefont{W.}~\bibnamefont{Zuo}},
  \bibinfo{journal}{Phys. Rev.} \textbf{\bibinfo{volume}{C 69}},
  \bibinfo{pages}{018801} (\bibinfo{year}{2004}).

\bibitem[{\citenamefont{Long et~al.}(2006{\natexlab{a}})\citenamefont{Long,
  Giai, and Meng}}]{Long:2006}
\bibinfo{author}{\bibfnamefont{W.~H.} \bibnamefont{Long}},
  \bibinfo{author}{\bibfnamefont{N.~V.} \bibnamefont{Giai}}, \bibnamefont{and}
  \bibinfo{author}{\bibfnamefont{J.}~\bibnamefont{Meng}},
  \bibinfo{journal}{Phys. Lett.} \textbf{\bibinfo{volume}{B 640}},
  \bibinfo{pages}{150} (\bibinfo{year}{2006}{\natexlab{a}}).

\bibitem[{\citenamefont{Long et~al.}(2007)\citenamefont{Long, Sagawa, Giai, and
  Meng}}]{Long:2007}
\bibinfo{author}{\bibfnamefont{W.~H.} \bibnamefont{Long}},
  \bibinfo{author}{\bibfnamefont{H.}~\bibnamefont{Sagawa}},
  \bibinfo{author}{\bibfnamefont{N.~V.} \bibnamefont{Giai}}, \bibnamefont{and}
  \bibinfo{author}{\bibfnamefont{J.}~\bibnamefont{Meng}},
  \bibinfo{journal}{Phys. Rev.} \textbf{\bibinfo{volume}{C 76}},
  \bibinfo{pages}{034314} (\bibinfo{year}{2007}).

\bibitem[{\citenamefont{Long et~al.}(2008)\citenamefont{Long, Sagawa, Meng, and
  Giai}}]{Long:2008}
\bibinfo{author}{\bibfnamefont{W.~H.} \bibnamefont{Long}},
  \bibinfo{author}{\bibfnamefont{H.}~\bibnamefont{Sagawa}},
  \bibinfo{author}{\bibfnamefont{J.}~\bibnamefont{Meng}}, \bibnamefont{and}
  \bibinfo{author}{\bibfnamefont{N.~V.} \bibnamefont{Giai}},
  \bibinfo{journal}{Europhysics Letters} \textbf{\bibinfo{volume}{82}},
  \bibinfo{pages}{12001} (\bibinfo{year}{2008}).

\bibitem[{\citenamefont{Liang et~al.}(2008)\citenamefont{Liang, VanGiai, and
  Meng}}]{Liang:2008}
\bibinfo{author}{\bibfnamefont{H.}~\bibnamefont{Liang}},
  \bibinfo{author}{\bibfnamefont{N.}~\bibnamefont{VanGiai}}, \bibnamefont{and}
  \bibinfo{author}{\bibfnamefont{J.}~\bibnamefont{Meng}},
  \bibinfo{journal}{Phys. Rev. Lett.} \textbf{\bibinfo{volume}{101}},
  \bibinfo{pages}{122502} (\bibinfo{year}{2008}).

\bibitem[{\citenamefont{Long et~al.}(2006{\natexlab{b}})\citenamefont{Long,
  Sagawa, Meng, and Giai}}]{Long:2006PS}
\bibinfo{author}{\bibfnamefont{W.~H.} \bibnamefont{Long}},
  \bibinfo{author}{\bibfnamefont{H.}~\bibnamefont{Sagawa}},
  \bibinfo{author}{\bibfnamefont{J.}~\bibnamefont{Meng}}, \bibnamefont{and}
  \bibinfo{author}{\bibfnamefont{N.~V.} \bibnamefont{Giai}},
  \bibinfo{journal}{Phys. Lett.} \textbf{\bibinfo{volume}{B 639}},
  \bibinfo{pages}{242} (\bibinfo{year}{2006}{\natexlab{b}}).

\bibitem[{\citenamefont{Bouyssy et~al.}(1987)\citenamefont{Bouyssy, Mathiot,
  VanGiai, and Marcos}}]{Bouyssy:1987}
\bibinfo{author}{\bibfnamefont{A.}~\bibnamefont{Bouyssy}},
  \bibinfo{author}{\bibfnamefont{J.~F.} \bibnamefont{Mathiot}},
  \bibinfo{author}{\bibfnamefont{N.}~\bibnamefont{VanGiai}}, \bibnamefont{and}
  \bibinfo{author}{\bibfnamefont{S.}~\bibnamefont{Marcos}},
  \bibinfo{journal}{Phys. Rev.} \textbf{\bibinfo{volume}{C 36}},
  \bibinfo{pages}{380} (\bibinfo{year}{1987}).

\bibitem[{\citenamefont{Baym et~al.}(1971{\natexlab{a}})\citenamefont{Baym,
  Pethick, and Sutherland}}]{Baym:1971AJ}
\bibinfo{author}{\bibfnamefont{G.}~\bibnamefont{Baym}},
  \bibinfo{author}{\bibfnamefont{C.~J.} \bibnamefont{Pethick}},
  \bibnamefont{and}
  \bibinfo{author}{\bibfnamefont{P.}~\bibnamefont{Sutherland}},
  \bibinfo{journal}{Astrophys. J.} \textbf{\bibinfo{volume}{170}},
  \bibinfo{pages}{299} (\bibinfo{year}{1971}{\natexlab{a}}).

\bibitem[{\citenamefont{Baym et~al.}(1971{\natexlab{b}})\citenamefont{Baym,
  Bethe, and Pethick}}]{Baym:1971NPA}
\bibinfo{author}{\bibfnamefont{G.}~\bibnamefont{Baym}},
  \bibinfo{author}{\bibfnamefont{H.~A.} \bibnamefont{Bethe}}, \bibnamefont{and}
  \bibinfo{author}{\bibfnamefont{C.~J.} \bibnamefont{Pethick}},
  \bibinfo{journal}{Nucl. Phys.} \textbf{\bibinfo{volume}{A 175}},
  \bibinfo{pages}{225} (\bibinfo{year}{1971}{\natexlab{b}}).

\bibitem[{\citenamefont{Reinhard et~al.}(1986)\citenamefont{Reinhard, Rufa,
  Maruhn, Greiner, and Friedrich}}]{Reinhard:1986}
\bibinfo{author}{\bibfnamefont{P.~G.} \bibnamefont{Reinhard}},
  \bibinfo{author}{\bibfnamefont{M.}~\bibnamefont{Rufa}},
  \bibinfo{author}{\bibfnamefont{J.}~\bibnamefont{Maruhn}},
  \bibinfo{author}{\bibfnamefont{W.}~\bibnamefont{Greiner}}, \bibnamefont{and}
  \bibinfo{author}{\bibfnamefont{J.}~\bibnamefont{Friedrich}},
  \bibinfo{journal}{Z. Phys.} \textbf{\bibinfo{volume}{A 323}},
  \bibinfo{pages}{13} (\bibinfo{year}{1986}).

\bibitem[{\citenamefont{Sharma et~al.}(1993)\citenamefont{Sharma, Nagarajan,
  and Ring}}]{Sharma:1993}
\bibinfo{author}{\bibfnamefont{M.~M.} \bibnamefont{Sharma}},
  \bibinfo{author}{\bibfnamefont{M.~A.} \bibnamefont{Nagarajan}},
  \bibnamefont{and} \bibinfo{author}{\bibfnamefont{P.}~\bibnamefont{Ring}},
  \bibinfo{journal}{Phys. Lett.} \textbf{\bibinfo{volume}{B 312}},
  \bibinfo{pages}{377} (\bibinfo{year}{1993}).

\bibitem[{\citenamefont{Bombaci and Lombardo}(1991)}]{Bombaci:1991}
\bibinfo{author}{\bibfnamefont{I.}~\bibnamefont{Bombaci}} \bibnamefont{and}
  \bibinfo{author}{\bibfnamefont{U.}~\bibnamefont{Lombardo}},
  \bibinfo{journal}{Phys. Rev.} \textbf{\bibinfo{volume}{C 44}},
  \bibinfo{pages}{1892} (\bibinfo{year}{1991}).

\bibitem[{\citenamefont{Lindblom}(1984)}]{Causality1}
\bibinfo{author}{\bibfnamefont{L.}~\bibnamefont{Lindblom}},
  \bibinfo{journal}{Astrophys. J.} \textbf{\bibinfo{volume}{278}},
  \bibinfo{pages}{364} (\bibinfo{year}{1984}).

\bibitem[{\citenamefont{Glendenning}(1992)}]{Glendenning:1992}
\bibinfo{author}{\bibfnamefont{N.~K.} \bibnamefont{Glendenning}},
  \bibinfo{journal}{Phys. Rev.} \textbf{\bibinfo{volume}{D 46}},
  \bibinfo{pages}{4161} (\bibinfo{year}{1992}).

\end{thebibliography}

\end{document}